\documentclass[a4paper,fleqn,usenatbib,useAMS]{mnras}

\usepackage{graphicx}	% Including figure files
\usepackage{amsmath}	% Advanced maths commands
\usepackage{amssymb}	% Extra maths symbols
\usepackage{multicol}        % Multi-column entries in tables
\usepackage{bm}		% Bold maths symbols, including upright Greek
\usepackage{pdflscape}	% Landscape pages
\usepackage[T1]{fontenc}
\usepackage{ae,aecompl}
\usepackage{mathptmx}
\title[Flares from fan-spine-like configuration]{Successive occurrences of quasi-circular ribbon flares in a fan-spine-like configuration involving hyperbolic flux tube}

\author[P. K. Mitra]{Prabir K. Mitra$^{1,2}$%
\thanks{Contact e-mail: \href{mailto:prair@prl.res.in}{prabir@prl.res.in}}%
and Bhuwan Joshi$^{1}$%
\\
$^{1}$Udaipur Solar Observatory, Physical Research Laboratory, Udaipur 313 001, India\\
$^{2}$Department of Physics, Gujarat University, Ahmedabad 380 009, India}

\pubyear{2020}

\begin{document}
\label{firstpage}
\pagerange{\pageref{firstpage}--\pageref{lastpage}}
\maketitle

\begin{abstract}
We present a comprehensive analysis of the formation and evolution of a fan-spine-like configuration that developed over a complex photospheric configuration where dispersed negative polarity regions were surrounded by positive polarity regions. This unique photospheric configuration, analogous to the geological ``atoll" shape, hosted four homologous flares within its boundary.  Computation of the degree of squashing factor ($Q$) maps clearly revealed an elongated region of high $Q$-values between the inner and outer spine-like lines, implying the presence of an hyperbolic flux tube (HFT). The coronal region associated with the photospheric atoll configuration was distinctly identified in the form of a diffused dome-shaped bright structure directly observed in EUV images. A filament channel resided near the boundary of the atoll region. The activation and eruption of flux ropes from the filament channel led to the onset of four eruptive homologous quasi-circular ribbon flares within an interval of $\approx$11 hours. During the interval of the four flares, we observed continuous decay and cancellation of negative polarity flux within the atoll region. Accordingly, the apparent length of the HFT gradually reduced to a null-point-like configuration before the fourth flare. Prior to each flare, we observed localised brightening beneath the filaments which, together with flux cancellation, provided support for the tether-cutting model of solar eruption. The analysis of magnetic decay index revealed favourable conditions for the eruption, once the pre-activated flux ropes attained the critical heights for torus instability.
\end{abstract}


\begingroup
\let\clearpage\relax
%\tableofcontents
\endgroup

%\newpage

\section{Introduction} \label{S-Intro}
Solar flares are sudden, localized enhancement of brightness in the solar atmosphere during which energy up to $\sim$10$^{32}$ erg can be released in the entire electromagnetic spectrum \citep[see review articles by][]{Fletcher2011, Benz2017}. It is well understood that magnetic field remains at the helm of all the catastrophic processes occurring in the solar atmosphere including flares, as the energy released during flares is supplied from the magnetic energy that is stored in the flaring region prior to the flare \citep[see e.g.,][]{Shibata2011}. Therefore, the pre-flare magnetic configuration plays a crucial role in determining the trigger and subsequent evolution of solar flares and associated eruptive phenomena \citep[e.g.,][]{Joshi2015, Joshi2017b, Hernandez2019b, Mitra2020a, Qiu2020}.

Traditionally, solar flares were observed to be associated with a pair of ribbon like brightening identified in chromospheric H$\alpha$ images, which were situated on the opposite sides of a polarity inversion line (PIL). To explain such parallel ribbon flares, a `standard flare model' was proposed combining the works of \citet{Carmichael1964, Sturrock1966, Hirayama1974, Kopp1976}, which is also known as the `CSHKP' model \citep{Shibata2011}. According to this model, magnetic reconnection takes place along a vertical current sheet, formed between the inflowing magnetic fields beneath an erupting prominence. During reconnection, magnetic energy gets transformed into heat and particle accelerations resulting in localised sudden flash in the solar atmosphere and highly accelerated electrons that are projected with almost relativistic speeds toward the lower, denser chromospheric layer of the Sun along the reconnected field lines. The accelerated electrons collide with the dense chromospheric plasma giving rise to hard X-ray (HXR) footpoint sources in association with EUV and optically observable conjugate ribbons termed as chromospheric flare ribbons \citep{Fletcher2013, Musset2015, Joshi2017, Kazachenko2017}. Despite the general success of the CSHKP model toward explaining the commonly observed features of eruptive parallel ribbon flares i.e., footpoint and looptop HXR sources, post-flare arcade, hot cusp etc., several studies have reported flaring activities to involve complex structures of flare ribbons and dynamics of overlying coronal loops implying that CSHKP model alone cannot explain all the flares \citep[e.g.,][]{Veronig2006, Joshi2009, Kushwaha2015, Mitra2018}.

It is also important to note that the CSHKP model, being a 2D model, can not explain the 3D aspects of typical solar flares such as, evolution of shear from the pre-flare loops to post-flare arcades; relative positions, shapes, and motions of the flare ribbons etc. To incorporate these features, the CSHKP model has been extended in 3D with numerical simulations \citep{Aulanier2012, Aulanier2013}. The 3D standard flare model suggests that small-scale current sheets are generated between the highly sheared pre-flare magnetic field configuration. Reconnection on these current sheets drives the transfer of differential magnetic shear, from the pre- to the post-eruptive configuration. With the evolution of the flare, as the eruption of the flux rope initiates, magnetic loops enveloping it straighten vertically and the current sheet extends along with them. Thus, magnetic reconnection continues beneath the erupting flux rope which is in line with the 2D standard flare model.

Further, the formation of the flux rope and the triggering of its eruption goes beyond the scope of the CSHKP model. Flux ropes are defined by a set of magnetic field lines that are wrapped around each other in a braided fashion or wrapped around a central axis \citep[braided and twisted flux ropes, respectively, see;][]{Prior2016}. Observationally, a flux rope can be identified in different forms: filament \citep{Zirin1988, Martin1998}, prominence \citep{Tandberg1995, Parenti2014}, coronal cavity \citep{Forland2013, Gibson2015}, hot channel \citep{Zhang2012, Cheng2013, Mitra2019, Sahu2020}, coronal sigmoid \citep{Rust1996, Manoharan1996, Joshi2017, Mitra2018} etc. The processes involved in the triggering of a flux rope from its stable condition and its successive evolution within the source region are rather complex and debatable \citep[see e.g.,][]{Chatterjee2013, Kumar2016, Prasad2020}. Different mechanisms have been proposed in this regard, which can be classified in two general groups: ideal instability and resistive instability. Ideal instability models put forward the idea that a flux rope can attain eruptive instability if the values of some parameters go beyond a critical value e.g., decay index ($n$; $n$=$-\frac{dlog(B_h)}{dlog(h)}$; $B_h$ and $h$ being horizontal magnetic field and height, respectively) more than 1.5 for torus instability \citep{Torok2006} or the twist of the flux rope more than $\approx$3.5$\pi$ for kink instability \citep{Torok2004}. On the other hand, resistive instability models recognize the role of initial small-scale reconnection in the active region as the triggering mechanism of a flux rope (or core field) e.g., initial reconnection beneath a flux rope or sheared arcade for the case of tether-cutting model \citep{Moore1992, Moore2001} or reconnection at a coronal null well above the core field for the case of breakout model \citep{Antiochos1999}. Once the MFR attains eruptive motion, magnetic reconnection initiates beneath the flux rope and two parallel ribbons are observed along with other observable flare signatures explained in the CSHKP model.

Morphologically, a completely different category of flares is circular (or quasi-circular) ribbon flares \citep{Masson2009} which are usually associated with a fan-spine configuration in a 3D null-point topology \citep{Lau1990, Sun2013}. Coronal null-points are locations in the solar corona, where the strengths of all the three components of magnetic field become locally zero \citep[see review by][]{Longcope2005}. Magnetic field beyond the immediate neighbourhood of the null-point is characterised by a spine line and a fan surface. Depending on the sign of the null-point (positive or negative), magnetic field lines approach the null-point along the spine line and move away from it along the fan surface \citep[for positive null; see Figure 4 in][]{Longcope2005}; or, approach the null along the fan surface and recede from it along the spine line (negative null). In the context of such 3D null-point configurations generating fan and spine lines, the anemone-type active regions where a compact magnetic region is surrounded by magnetic regions of opposite polarity \citep[see,][]{Shibata1994} are of special significance. The inner compact region is connected with the surrounding opposite polarity region by small-scale closed magnetic loops while a set of relatively large field lines connect the surrounding polarity to a remote region of polarity similar to that of the inner compact region. In this way, the two sets of field lines constitute two sets of fan lines (inner and outer fan lines) and two sets of spine lines (inner and outer spine lines). The two sets of fan lines are separated by a dome-shaped surface (i.e., fan separatrix) characterised by high degree of squashing factor \citep[$Q$; see,][]{Priest1995, Titov2002}, which intersects the spine lines at the null-point \citep{Sun2013}. In general, domains corresponding to drastic changes in the magnetic field connectivity gradient are identified as quasi-separatrix layers \citep[QSLs; see e.g.,][]{Janvier2013}. While the values of $Q$ corresponding to QSLs are high \citep[$\gg$2; see,][]{Aulanier2005}, null-points can be characterised by $Q\rightarrow\infty$. The finite values of $Q$ at the QSLs imply that although magnetic fields show drastic change in the connectivity they are still continuous, which is contrary to the cases of null-points where magnetic field becomes discontinuous. In complex photospheric configurations, e.g., those formed by two bipolar sunspots, a pair of photospheric null-points of opposite signs may exist which are connected by `separators' \citep{Titov2002}. A separator can be identified by narrow elongated strips of high $Q$-values, i.e. a QSL, with a pair of null-points at both of its ends. Further, a generalisation of the concept of separator lines reveals a special geometrical feature called `Hyperbolic Flux Tubes' \citep[HFTs;][]{Titov2002} which can be understood as the intersection of two QSLs. The middle of an HFT is characterised by `X'-type cross section comprised of high $Q$-values. Such structures of high $Q$-values are preferred sites for the formation of current sheets and initiation of magnetic reconnection \citep{Titov2003}. Moreover, magnetic field lines can constantly change their connectivities along the QSLs as a consequence of local diffusion in the region, allowing neighboring field lines to exchange connectivities \citep{Aulanier2006}. This can be observed as an apparent slipping or flipping motion of loop connectivities and are termed as `slipping reconnction' \citep[see e.g.,][]{Priest1995, Aulanier2006, Janvier2013}.

With the advancements of observational facilities and numerical techniques, in the recent years a number of studies have reported flaring activities that involved both the circular and parallel ribbons where the parallel ribbons usually reside at the inside edge of the circular ribbon \citep[e.g.,][]{Joshi2015, JoshiN2017, Hernandez2017,LI2017,  Xu2017, LiH2018, Li2018, Hou2019, Shen2019, Devi2020}. Such events usually develop as a small flux rope erupts within a fan-dome which then triggers reconnection at the null-point. The null-point reconnection itself is a complex, multi-stage mechanism which initially includes slipping reconnection at the quasi-separatrix surface at the fan-dome giving rise to the circular or quasi-circular ribbon and during the subsequent stage, interchange reconnection takes place between the inner close fan lines and the outer open spine lines causing the remote brightening \citep[see,][]{Masson2009}. In response to the interchange reconnection at the coronal null-point, collimated ejection of plasma i.e., coronal jets or H$\alpha$-surges have been identified in several studies \citep{Pariat2009b, Pariat2010}. These findings point toward the fact that magnetic configurations on the Sun could be very complex and more studies are extremely essential toward reaching at a general understanding of the complex sunspot configurations and associated flaring activities.

In this article, we report four homologous quasi-circular ribbon flares from the active region NOAA 11977, which were triggered by erupting filaments from the circular ribbon region. With the help of high resolution images of Atmospheric Imaging Assembly \cite[AIA;][]{Lemen2012} and Helioseismic and Magnetic Imager \citep[HMI;][]{Schou2012} on board the \textit{Solar Dynamics Observatory} \citep[\textit{SDO};][]{Pesnell2012}, we study the evolution of the active region and the complex flares in detail. Magnetic field modelling based on a `Non-linear Force Free Field' (NLFFF) method has revealed a fan-spine-like configuration associated with the flaring region. The most important finding of this study is the absence of coronal null-point in the fan-spine-like configuration. Instead, the calculation of $Q$ revealed the presence of an HFT between the inner and outer spine-like lines. In Section \ref{Obs_Data}, we provide a brief description of the observational data sources and the image analysis techniques along with the numerical methods used in this article. We discuss the morphology and evolution of the active region as well as give a brief account of all the flares produced by it in Section \ref{sec_ar}. Results obtained from imaging analysis of the two circular ribbon flares and NLFFF extrapolation are presented in Sections \ref{sec_all_flares} and \ref{extrapolation}. We discuss and interpret the results in Section \ref{dscsn}.

\section{Observational Data and Analysis Techniques} \label{Obs_Data}
For EUV imaging, we have utilised the 12 s cadence, 4096$\times$4096 pixel full disk observations from the AIA on board the \textit{SDO} with pixel resolution of 0$\farcs$6. For the chromospheric imaging of the Sun, we have used the 2048$\times$2048 pixel full disk images in the H$\alpha$ passband with a pixel resolution of $\approx$1$\farcs$0, obtained from the archive Global Oscillation Network Group \citep[GONG;][]{Harvey1996, harvey2011}.  We have studied the photospheric structures associated with the active region NOAA 11977 by using the 45 s cadence, 4096$\times$4096 pixel full disk continuum and line-of-sight (LOS) magnetogram observations with spatial sampling resolution of 0$\farcs$5 pixel$^{-1}$ by HMI on board the \textit{SDO}. The HMI LOS intensity and magnetogram images were further processed with the IDL-based algorithm `\textit{hmi\_prep}' to co-align them with AIA pixel resolution. Coronal magnetic fields were extrapolated by employing the optimisation based Non-Linear Force Free Field (NLFFF) extrapolation method developed by \citet{Wiegelmann2010, Wiegelmann2012}. For the purpose, we have used the vector magnetogram data from the `\textit{hmi.sharp$\_$cea$\_$720s}' series of HMI at a reduced spatial resolution of 1$\farcs0$ pixel$^{-1}$ as the input boundary condition. Extrapolations were done within a volume of dimensions 453$\times$270$\times$240 pixels which corresponds to the physical dimension of $\approx$328$\times$196$\times$174 Mm$^3$. Based on the NLFFF extrapolation results, we calculated the degree of squashing factor ($Q$) and twist number ($T_w$) in the extrapolation volume by using the IDL-based code developed by \citet{Liu2016}. In order to locate 3D null-points within the extrapolation volume, we used the trilinear method as suggested by \citet{Haynes2007}. For the purpose, the whole active region volume was divided into grid cells of dimension 2$\times$2$\times$2 pixels. The first step of the trilinear method is to quickly scan through every grid cell by examining the signs of each component of magnetic field at all the eight corners of the grid cells. If any of the three components have same sign at all the eight corners, a null-point can not reside within the grid cell and therefore, the corresponding cell is excluded from further analysis. Each of the remaining other cells is then further divided into 100$\times$100$\times$100 sub-grid cells and the threshold $\bigtriangleup x \leqslant 2$ sub-grid cell-width was used for locating null-points. For visualizing the modelled field lines and the distribution of $Q$ in the active region volume, we have used the Visualization and Analysis Platform for Ocean, Atmosphere, and Solar Researchers \citep[VAPOR\footnote{\url{https://www.vapor.ucar.edu/}};][]{Clyne2007} software.

\section{Structure and Evolution of the active region NOAA 11977} \label{sec_ar}
The active region NOAA 11977 appeared on the eastern limb of the Sun on 2014 February 11 as a simple $\alpha$-type active region. It quickly transformed into a relatively more complex $\beta$-type on the very next day. The active region gradually developed into $\beta\gamma$-type on 2014 February 14 and remained so for the next four days. Notably, the active region started to decay in its area since 2014 February 15. On 2014 February 16, an intriguing configuration of magnetic fields, involving complex distribution and topology, developed in the westernmost part of the active region which we study comprehensively in Section \ref{atoll}. Notably, one M and three C-class flares originated from this region within an interval of $\approx$11 hours on February 16 (see Table \ref{table1}). The magnetic complexity of the active region reduced to $\beta$-type on 2014 February 19. The active region disappeared from the western limb of the Sun on 2014 February 23 as an $\alpha$-type sunspot.

\begin{figure*}

\includegraphics[width=\textwidth]{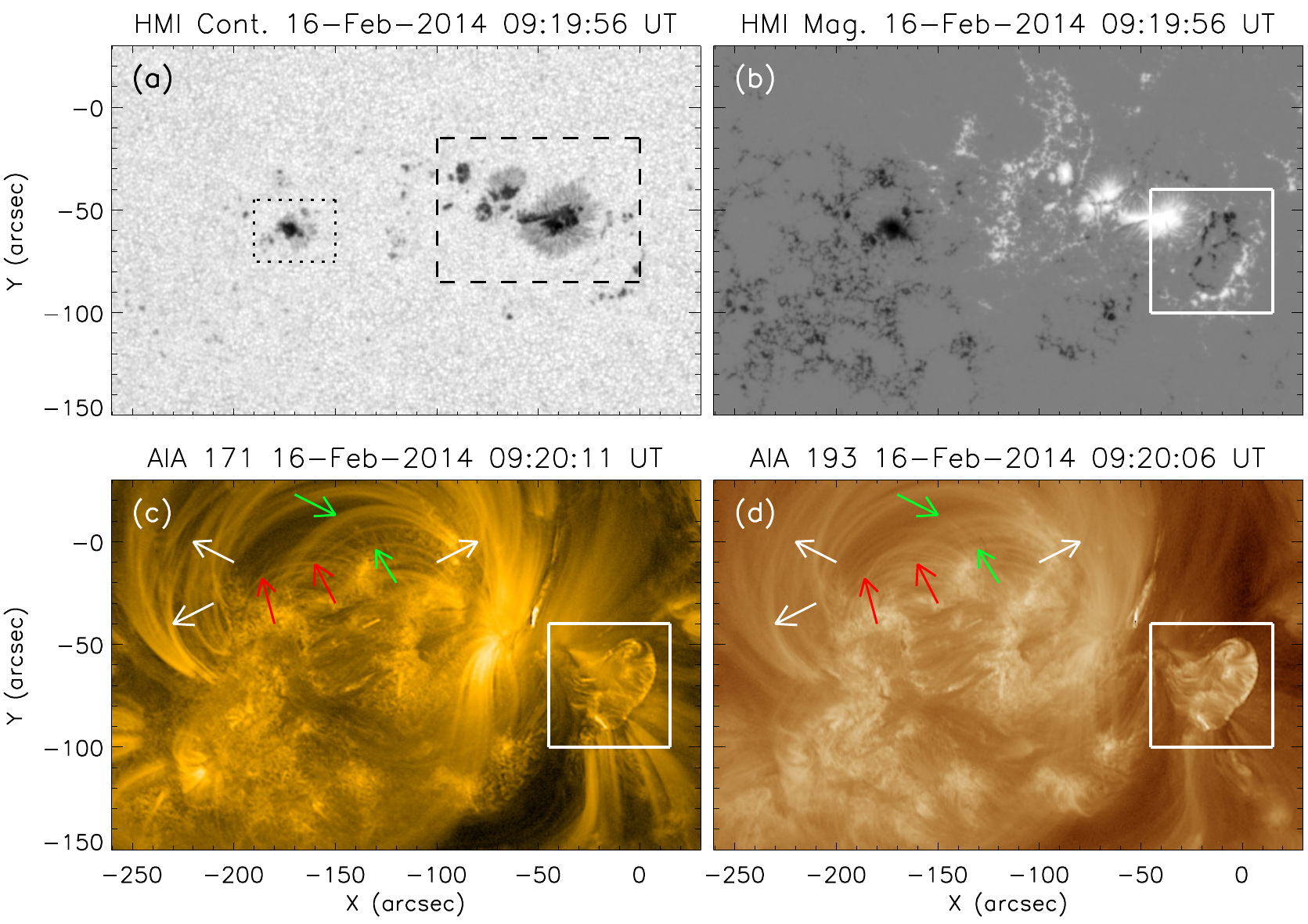}
\caption{The morphology of the active region NOAA 11977 on the photosphere (panels (a) and (b)) and different coronal temperatures (panels (c) and (d)). The flaring activity occurred from the region shown within the white box in panels (b)--(d).}
\label{ar_overview}
\end{figure*}

\subsection{Morphology of the active region NOAA 11977} \label{sec_AR_morph}
In Figure \ref{ar_overview}, we show a comparison of the photospheric structure of the active region with its coronal configuration, prior to the M-class flare on 2014 February 16. We find that, the active region was comprised of a few prominent sunspots and many pores. We consider two subregions of the active region: the leading sunspot group and the trailing sunspot group (shown by the dashed and dotted boxes in Figure \ref{ar_overview}(a), respectively). Comparison of a co-temporal LOS magnetogram (Figure \ref{ar_overview}(b)) of the active region with the intensity image reveals that the leading sunspot group was consisted of mostly positive polarity while the trailing part of the active region was dominated by negative polarity magnetic field. However, the most interesting aspect of the active region, in the context of our analysis, is the magnetic configuration that developed in the extreme western part of it where magnetic patches of positive and negative polarities formed a configuration similar to an `atoll' (within the white box in Figure \ref{ar_overview}(b)). AIA images suggest that the active region, on the whole, consisted of different sets of coronal loops of varying spatial extents and projected heights. In Figures \ref{ar_overview}(c) and (d), we recognize some of these loops by arrows of different colors (white, green and red).

We note striking coronal structures over the photospheric atoll region which can be easily distinguished by its peculiar morphology. The region is outlined by the white boxes in Figures \ref{ar_overview}(c) and (d). The EUV images of the coronal region, over the atoll-shaped magnetic structure in the photosphere, clearly reveal enhanced structured brightening within a sharp boundary that forms an oval-shaped feature. All the four events of 2014 February 16 occurred within this region.

\subsection{Formation of the `magnetic atoll' region} \label{atoll}
\begin{figure*}
\includegraphics[width=0.99\textwidth]{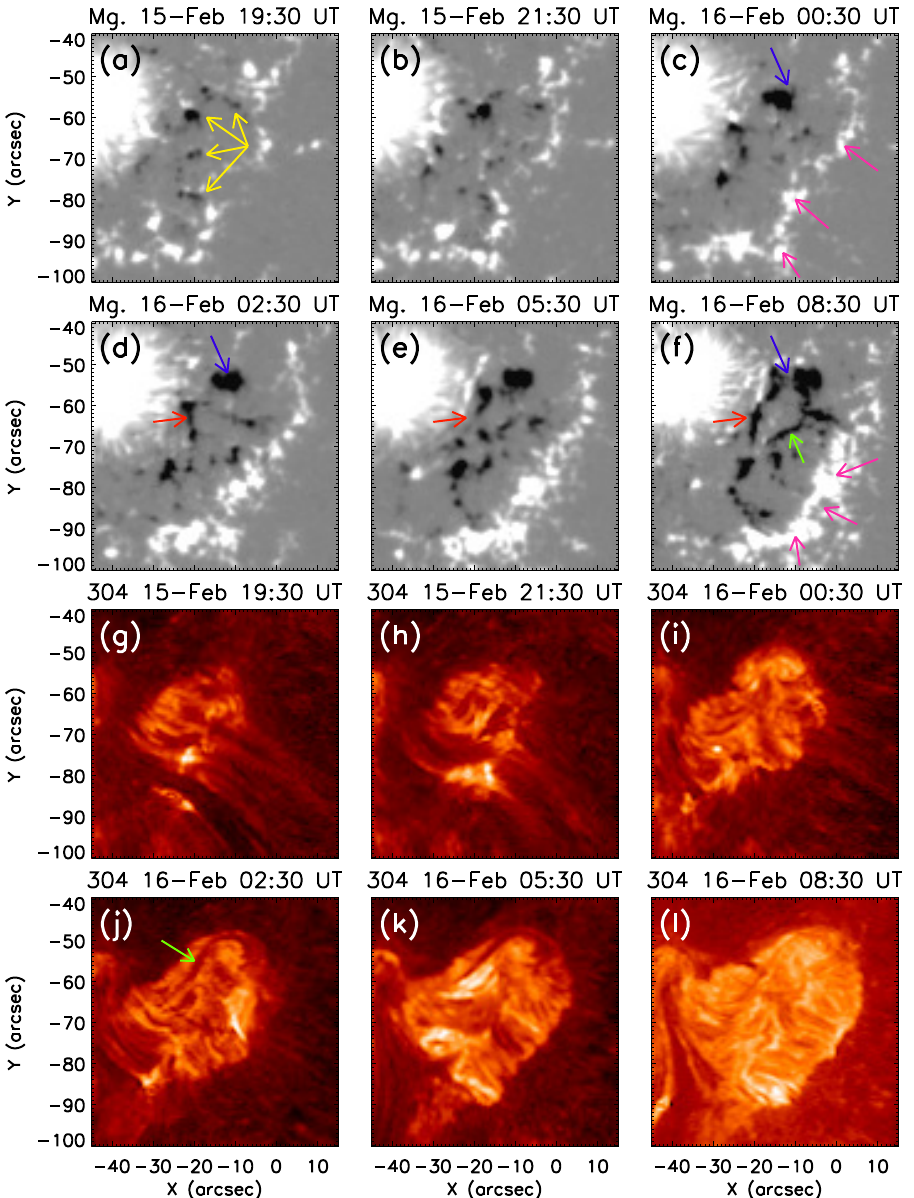}
\caption{Series of HMI LOS magnetograms showing the development of the magnetic `atoll' region (panels (a)--(f)). Arrows of different color indicate few prominent locations of negative magnetic flux emergence. Panels (g)--(l) are AIA 304 \AA\ images associated with the magnetic atoll region displaying the formation of the diffused circular brightening region. All the images are derotated to 2014 February 16 09:20 UT.}
\label{dome_form}
\end{figure*}

In order to understand the formation and development of the magnetic atoll region, in Figures \ref{dome_form}(a)--(f), we present a series of LOS magnetograms corresponding to the photospheric region (shown within the white box in Figure \ref{ar_overview}(b)). Since, the diffused brightening region was intrinsically related with the magnetic atoll region, we have also plotted co-temporal AIA 304 \AA\ images of the same region in Figures \ref{dome_form}(g)--(l). Our observations reveal that the atoll region started to develop from 2014 February 15 with the emergence a of few small negative polarity flux regions. In Figure \ref{dome_form}(a), we indicate these emerging negative polarity regions by the yellow arrows. With time, negative polarity magnetic flux kept emerging in this region and by the first hour of 2014 February 16, a relatively large magnetic patch of negative polarity was developed (indicated by the blue arrow in Figure \ref{dome_form}(c)). Co-temporal AIA 304 \AA\ images suggest that the distinct oval-shaped region displaying enhanced-diffused brightening started to build-up from around this time onward (cf. Figures \ref{dome_form}(i) and (c)). Gradually, this diffused brightening region became more prominent and extended spatially (cf. Figures \ref{dome_form}(i)--(l)). In the same duration, positive flux also emerged in this region which surrounded the negative polarity patches in the south-west direction (cf. the regions indicated by the pink arrows in Figures \ref{dome_form}(c) and (f)). In general, the atoll region can be characterised by newly emerged and spatially dispersed prominent patches of negative magnetic flux surrounded by positive polarity regions from two sides i.e., northeast (by the positive polarity sunspot) and southwest (by the dispersed positive polarity patches). Further, from HMI LOS magnetograms during the early hours of 2014 February 16, we could identify several instances of flux emergence and cancellation in the magnetic atoll region. We have identified a few events of flux changes by blue, red and green arrows in Figures \ref{dome_form}(c)--(f). During the same interval, we also observed formation of a filament channel in the diffused brightening region along the PIL between the positive polarity sunspot and adjacent negative polarity patches. We have indicated it by a green arrow in Figure \ref{dome_form}(j).

\section{Four successive flares from the magnetic atoll region} \label{sec_all_flares}
\begin{figure*}
\includegraphics[width=\textwidth]{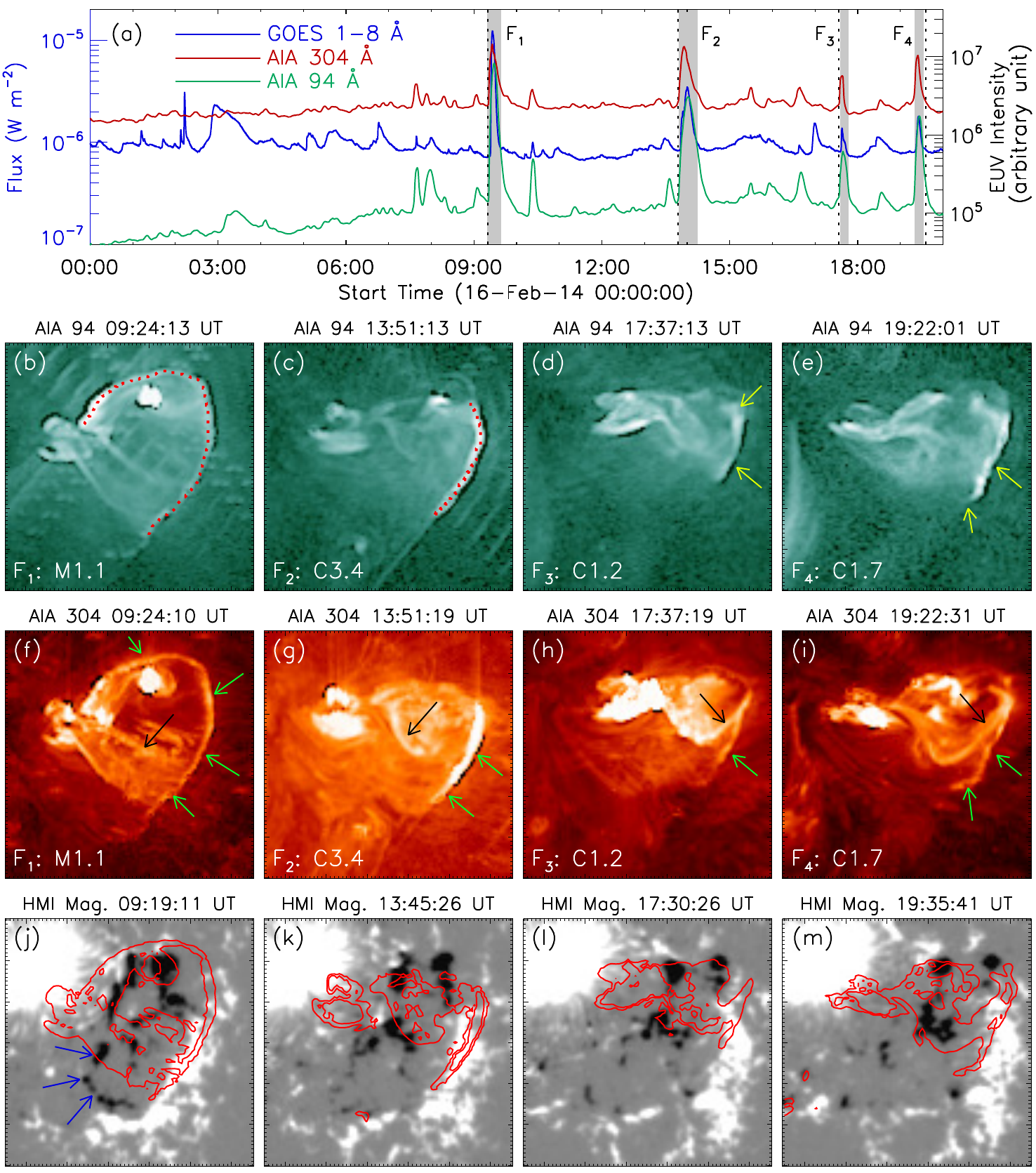}
\caption{Panel (a): GOES 1--8 \AA\ flux (blue curve), AIA 304 \AA\ (red curve) and AIA 94 \AA\ (green curve) intensities showing the onset and evolution of four flares from the diffused brightening region. The duration of the flares are indicated by four shaded regions. The individual fares are named as `F$_1$', `F$_2$', `F$_3$' and `F$_4$'. Panels (b)--(e): AIA 94 \AA\ images of the diffused brightening region during the peak phases of the four flares. The red dotted curves in panels (b) and (c) indicate the quasi-circular ribbon observed during F$_1$ and F$_2$ while the yellow arrows in panels (d)--(e) indicate arc-shaped ribbons during F$_3$ and F$_4$ flares, respectively. Panels (j)--(i): AIA 304 \AA\ images showing the peak phases of the four flares. The green arrows in different panels indicate the quasi-circular (panel (f)), semi-circular (panel (g)) and arc-shaped (panels (h) and (i)) ribbons observed during the flares. The black arrows indicate the erupting filament. Panels (j)--(m): HMI LOS magnetograms of the `magnetic atoll' region prior to the onset of the F$_1$--F$_3$ flares (panels (j)--(l)) and after the F$_4$ flare (panel (m)). The blue arrows in panel (j) indicate few negative flux region which disappeared with time. Contours of co-temporal EUV intensities in AIA 304 \AA\ are overplotted on the bottom row. Contour levels are [4, 80]\%, [12, 60]\%, [11, 60]\% and [9, 40]\% of the corresponding peak intensities in panels (j), (k), (l) and (m), respectively. An animation associated with this figure is attached in the online supplementary materials.}
\label{allflares}
\end{figure*}

The magnetic atoll region produced four successive flares on 2014 February 16. In Figure \ref{allflares}(a), we plot SXR flux in the GOES 1--8 \AA\ channel along with EUV intensities of AIA 94 and 304 \AA\ channels from 2014 February 16 00:00 UT--20:00 UT which displays the onset and temporal evolutions of the four flares (see summary of events in Table \ref{table1}). These multi-wavelength lightcurves suggest that the first flare of GOES class M1.1 initiated at $\approx$09:20 UT. This short lived flare reached to its peak at $\approx$09:26 UT. The second flare (GOES class C3.4) initiated at $\approx$13:48 UT and after a relatively extended rise phase of $\approx$12 minutes, reached its peak intensity. Interestingly, the subsequent two flares originating from this region were rather impulsive and short lived. According to the GOES 1--8 \AA\ flux profile, these C class flares originated at $\approx$17:35 UT and 19:20 UT while their impulsive phases lasted for only $\approx$3 minutes each. In Figure \ref{allflares}(a), we indicate the durations of the four flares by gray-shaded intervals and assign the notations `F$_1$', `F$_2$', `F$_3$' and `F$_4$' to the flares in the chronological order.

\begin{table*}
\centering
\caption{Summary of the flares occurred on 2014 February 16 from the magnetic atoll region of NOAA 11977}
\label{table1}
\begin{tabular}{ccccc}
\hline
\hline
Flare&Location&Flare timing (UT)&GOES class&Remarks\\
Id.&&Start/ Peak/ End&\\
\hline
F$_1$&S10E00&09:20/ 09:26/ 09:29&M1.1&Observation of quasi-circular ribbon\\
F$_2$&S10W01&13:48/ 14:00/ 14:05&C3.4&Observation of shortened quasi-circular ribbon\\
F$_3$&S10W03&17:35/ 17:38/ 17:40&C1.2&Observation of arc-shaped ribbon\\
F$_4$&S10W04&19:20/ 19:23/ 19:28&C1.7&Observation of arc-shaped ribbon\\
\hline
\end{tabular}
\end{table*}

In Figures \ref{allflares}(b)--(e) and (f)--(i), we show AIA 94 \AA\ and AIA 304 \AA\ images, respectively, of the diffused brightening region during the peak phases of the circular ribbon flares. From these images it becomes evident that, all the four flares were associated with some degree of ribbon like brightening along the circumference of the diffused brightening region. We observed quite prominent signatures of an extended quasi-circular ribbon brightening during F$_1$, which is delineated by the red dotted curve in Figure \ref{allflares}(b) and the green arrows in Figure \ref{allflares}(f). The quasi-circular ribbon shortened significantly during F$_2$ which we have indicated by the red dotted curve in Figure \ref{allflares}(c) and the green arrows in Figure \ref{allflares}(g). During F$_3$ and F$_4$ the flare ribbons along with only a small portion of the boundary of the diffused brightening region displayed enhanced emission. This arc-shaped flare brightening is indicated by the yellow arrows in Figures \ref{allflares}(d)--(e) and the green arrows in Figures \ref{allflares}(h)--(i).

In Figures \ref{allflares}(j)--(m), we plot preflare HMI LOS magnetograms corresponding to all the four flares, respectively. We find that the negative polarity flux regions from the magnetic atoll region continuously decayed with the evolution of each flares. This cancellation of negative flux was much more prominent in the southern part of the atoll region than the northern part (cf. the negative flux regions indicated by blue arrows in Figures \ref{allflares}(j) and (m)). For a better understanding of the relation between the photospheric flux regions and the flare emission, we plotted contours of EUV intensities in AIA 304 \AA\ channel over the LOS magnetograms in Figures \ref{allflares}(j)--(m). From the overplotted contours, it becomes clear that the quasi-circular ribbon brightening during the peak phases of the flares were co-spatial with the positive polarity flux regions situated at the boundary of the photospheric atoll region.

\begin{figure}
\includegraphics[width=0.35\textwidth]{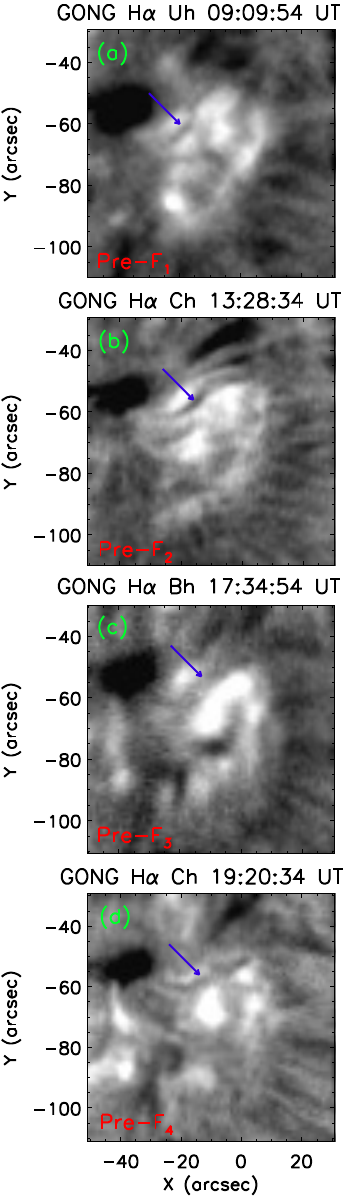}
\caption{Selective of H$\alpha$ images of the diffused brightening region during the pre-flare phases of the quasi-circular ribbon flares, recorded by different stations of the GONG network. The corresponding GONG stations are indicated in the titles in each panel as Uh (Udaipur Solar Observatory, India), Ch (Cerro Tololo Inter-American Observatory, Chile) and Bh (Big Bear Solar Observatory, USA). The blue arrows indicate the presence of small filaments prior to the flares.}
\label{halpha}
\end{figure}

In Figure \ref{halpha}, we plot chromospheric images of the diffused brightening region in the H$\alpha$ passband during the pre-flare phases of each flares. From these images, we clearly identified impressions of the filaments, implying recursive development of filaments at the filament channel situated along the PIL between the positive polarity sunspot and the parasitic negative polarity patches (Figure \ref{dome_form} and Section \ref{atoll}). It is worth mentioning that the evolution of the quasi-circular ribbon flares in the H$\alpha$-passband (not shown here) was broadly similar to that observed in EUV channels. 

\begin{figure*}
\includegraphics[width=0.85\textwidth]{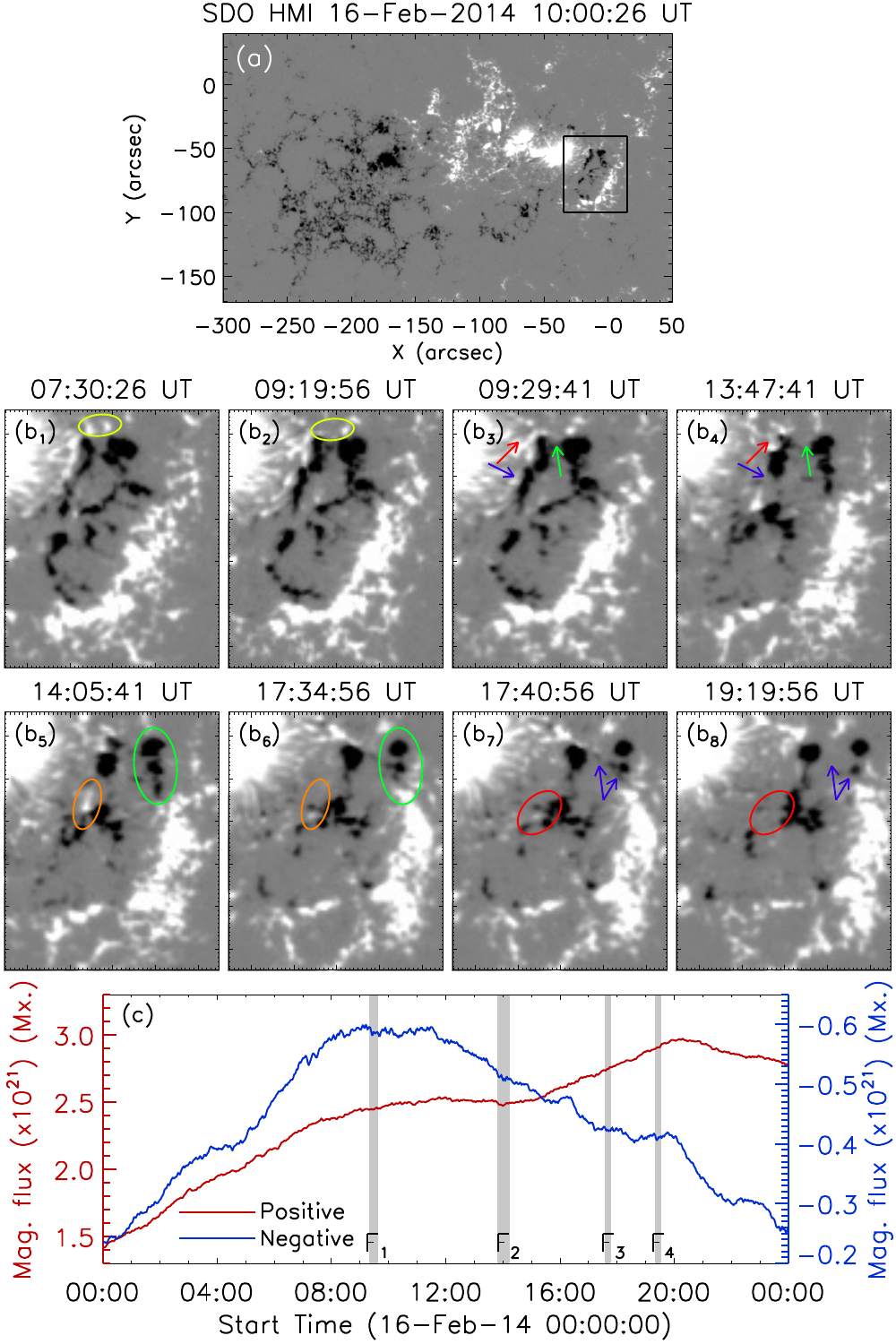}
\caption{Panel (a): LOS magnetogram of the active region NOAA 11977 where the magnetic atoll region is enclosed within the black box. Panels (b$_\textrm{1}$)--(b$_\textrm{8}$): Selective LOS magnetograms of the atoll region displaying several instances of flux cancellation. We have indicated few of such flux cancellation by arrows and ovals of different colours in different panels. Panel (c): Evolution of magnetic flux in the magnetic atoll region indicated in panel (a). The shaded intervals in panel (c) indicate the durations of flares occurred from the magnetic atoll region.}
\label{ar_flux}
\end{figure*}

In order to have a comprehensive understanding towards the influence of photospheric flux on the evolution of the diffused brightening region as well as on the homologous flares originated from it, in Figure \ref{ar_flux}, we examine the evolution of photospheric flux over a prolonged period that also covers the time-span of the four flares. For the purpose, we emphasise on the small-scale flux changes within the atoll region (in Figures \ref{ar_flux}(b$_1$)--(b$_8$)) and the temporal evolution of flux from it (Figure \ref{ar_flux}(c)) by analysing HMI LOS magnetograms. The atoll region is enclosed in Figure \ref{ar_flux}(a) by the black box. We clearly identified several instances of small-scale flux cancellation; a few of them are indicated by different colours of ovals and arrows in Figures \ref{ar_flux}(b$_1$)--(b$_8$). During the pre-flare phase of F$_1$ (Figures \ref{ar_flux}(b$_1$)--(b$_2$)) we observed cancellation of positive flux at the region of the PIL of the filament, which is indicated by the yellow ovals. During the interval between F$_1$ and F$_2$, we could identify multiple instances of flux cancellation of both polarities at the PIL region situated at the northern part of the atoll region (cf. the arrow heads of the red, blue and green arrows in Figures \ref{ar_flux}(b$_3$)--(b$_4$)). Between F$_3$ and F$_4$, a major part of a relatively prominent negative flux region got decayed (cf. the region enclosed by the green ovals in Figures \ref{ar_flux}(b$_5$)--(b$_6$)). We further observed cancellation of positive flux in the atoll region which we have highlighted within the orange ovals in Figures \ref{ar_flux}(b$_5$)--(b$_6$). This region was observed with flux cancellation between F$_3$ and F$_4$ also which is indicated by the red arrows in Figures \ref{ar_flux}(b$_7$)--(b$_8$). We also observed cancellation of small negative fluxes from the atoll region in this duration which are indicated by the blue arrows in Figures \ref{ar_flux}(b$_7$)--(b$_8$).

In general, the atoll region experienced a significant decrease of negative flux (Figures \ref{allflares}(j)--(m) and \ref{ar_flux}(b$_1$)--(b$_8$)) which is readily recognised by the temporal evolution of negative flux (blue curve in Figure \ref{ar_flux}(c)). The flux profiles suggest that in the first part of 2014 February 16, the flaring region (enclosed by the black box in Figure \ref{ar_flux}(a)), underwent a rapid increase of negative flux beside relatively moderate positive flux enhancement (see 00:00 UT -- 08:00 UT in Figure \ref{ar_flux}(c)). Notably, this period can be characterised by the formation of the atoll and associated diffused brightening region (Section \ref{atoll} and Figure \ref{dome_form}). The first flare (F$_1$) initiated when negative flux reached to its peak; afterwards negative flux from the region remained relatively unchanged for $\approx$3 hours and then continuously decayed till $\approx$19:30 UT. After this time, negative flux maintained an approximately constant level for $\approx$1.5 hours when the last flare from this region (F$_4$) took place. From $\approx$20:00 UT, negative flux primarily decayed characterising the decay of the overall diffused brightening region. On the other hand, positive flux in this region displayed stepwise enhancement till $\approx$20:00 UT after which decayed slowly till end end of the period of our calculation.

\subsection{Morphological evolution of quasi-circular ribbon flares} \label{sec_flares}
All the four flares originating from the oval-shaped diffused brightening region initiated with the activation and eruption of the filaments situated along the PIL between the positive polarity sunspot the negative flux regions. We recall that the coronal region showing oval-shaped diffused brightening was lying over the peculiar magnetic atoll region at the photosphere (see Section \ref{atoll} and Figure \ref{dome_form}). Further, during all the four flares, we observed flare ribbons along the boundary of the diffused brightening region which forms quasi-circular ribbons (Figure \ref{allflares}). The quasi-circular ribbon was the most prominent and extended during F$_1$ of GOES class M1.1. The subsequent three flares were very similar in their onset and evolution. However, as explained in Section \ref{sec_all_flares} (Figure \ref{allflares}), the length of quasi-circular ribbon decreased progressively during the subsequent flares (i.e., F$_2$--F$_4$). More importantly, during F$_2$--F$_4$, the filament evolved with a complete blow-out type eruption from the diffused brightening region while such catastrophic eruption of the filament was not observed during F$_1$. In this section, we focus on the detailed morphological evolution of F$_1$ and F$_2$ flares.

\subsubsection{Filament eruption and quasi-circular ribbon during F$_1$} \label{sec_M1.1}
\begin{figure*}
\includegraphics[width=0.93\textwidth]{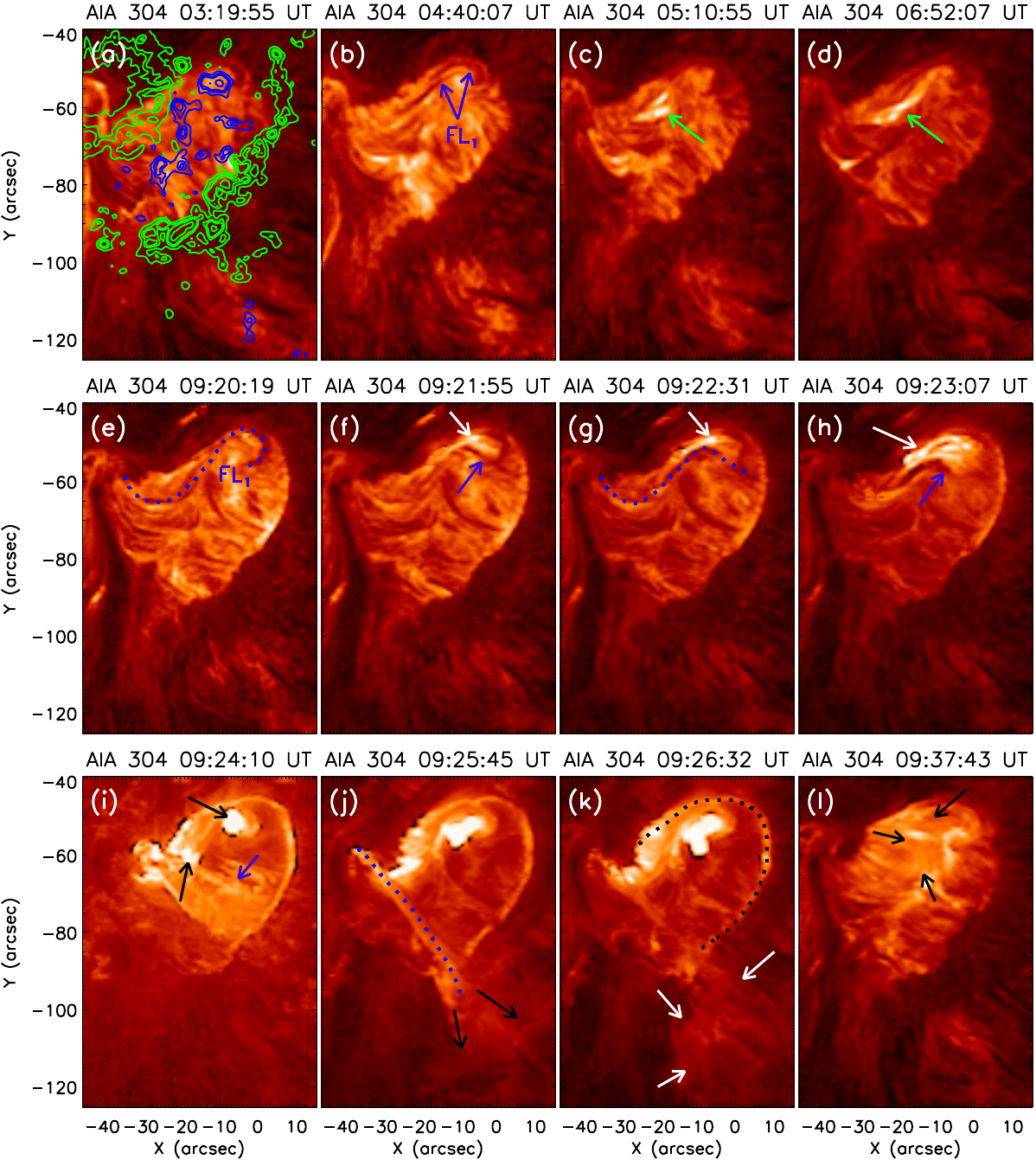}
\caption{Representative AIA 304 \AA\ images depicting filament eruption and associated M1.1 flare. The two blue arrows indicate a filament which is referred to as `FL$_1$'. The green arrows in panels (c) and (d) indicate small-scale brightenings observed at the location of the filament. In panel (e), we outline the `S' shaped filament by dotted curve. The filament during its initial eruption phase is shown by blue arrows in panels (f) and (h) and blue dotted line in panel (g). The white arrows in panels (c)--(h) indicate initial flare brightening beneath the northern end of the erupting filament. The final stage of filament eruption from the active region is outlined by the blue arrow in panel 9i) and blue dotted curve in panel (j). The black arrows in panel (e) indicate parallel flare brightening during the impulsive phase of the flare. The black arrows in panel (j) and white arrows in panel (k) indicate erupting plasma. The black dotted curve in panel (k) indicate the quasi-circular ribbon. Co-temporal LOS HMI magnetogram contours at levels $\pm$[300, 500, 1000] G are plotted in panel (a). Green and blue contours refer to positive and negative polarity, respectively. An animation associated with this figure is attached in the online supplementary materials.}
\label{M1.1}
\end{figure*}

In Figure \ref{M1.1}, we plot a series of AIA 304 \AA\ images of the diffused brightening region to investigate the evolution and eruption of the filament in association with F$_1$ i.e., the GOES M1.1 flare. Our observations suggest that the impression of the filament at the northern boundary of the diffused brightening region was first identified at around $\approx$02:30 UT (Section \ref{atoll} and Figure \ref{dome_form}). In Figure \ref{M1.1}(b), we indicate the initial filament by the blue arrows and we will refer this filament as FL$_1$ henceforth. A comparison of the location of FL$_1$ with the LOS magnetograms in Figure \ref{M1.1}(a) immediately reveals that the filament was situated along the PIL between the positive polarity sunspot and the dispersed negative magnetic field regions. The filament was associated with a series of localised small-scale brightenings (indicated by the green arrows in Figures \ref{M1.1}(c) and (d)) during an extended period prior to the onset of the GOES M-class flare (i.e, F$_1$). The filament was observed most prominently at $\approx$09:20 UT which is indicated by the blue dotted curve in Figure \ref{M1.1}(e). After $\approx$09:21 UT, a subtle enhancement in the brightness at the northern leg of the filament was observed which was immediately followed by eruption of the filament from its northern end. In Figure \ref{M1.1}(f), we indicate the erupting part of the filament by a blue arrow and the localized brightening by a white arrow. The enhanced brightening advanced southward (cf. the white arrowheads in Figures \ref{M1.1}(d) and (c)) as the erupting front of the filament induced upward erupting motion in the southern part of the filament also. The southward induced eruption of the filament is further outlined by the blue dotted curve in Figure \ref{M1.1}(g) and indicated by the blue arrows in Figures \ref{M1.1}(h) and (i). Two localised ribbon-like bright structures formed at both sides of the filament (indicated by the black arrows in Figure \ref{M1.1}(i)) at around 09:24 UT, which resembles with `standard flare ribbons'. At the same time, a circular ribbon brightening was prominently observed at the boundary of the diffused brightening region. After $\approx$09:25 UT, the filamentary structure (FL$_1$) appeared as a straight, long structure with one end still attached to its initial location (outlined by the blue dotted curve in Figure \ref{M1.1}(j)). Interestingly, during this time, it was observed as a bright structure suggesting strong heating of the filamentary materials during the flaring process. The open end of the filamentary structure was associated with plasma eruption after $\approx$09:25 UT. The direction of the erupting plasma is indicated by the arrows in Figure \ref{M1.1}(j). Notably, both the parallel and circular ribbon brightening increased up to $\approx$09:27 UT (the circular ribbon brightening is outlined by the black dotted curve in Figure \ref{M1.1}(g)) after which, flare brightenings from the active region slowly decreased while eruption continued. In Figures \ref{M1.1}(k), we indicate the erupting plasma by white arrows. Formation of post-flare arcade was observed after $\approx$09:32 UT which sustained till $\approx$09:42 UT. We have indicated the post-flare arcade by black arrows in Figure \ref{M1.1}(l).

\subsubsection{Filament eruption and shortened quasi-circular ribbon during F$_2$} \label{sec_C3.4}
\begin{figure*}
\includegraphics[width=\textwidth]{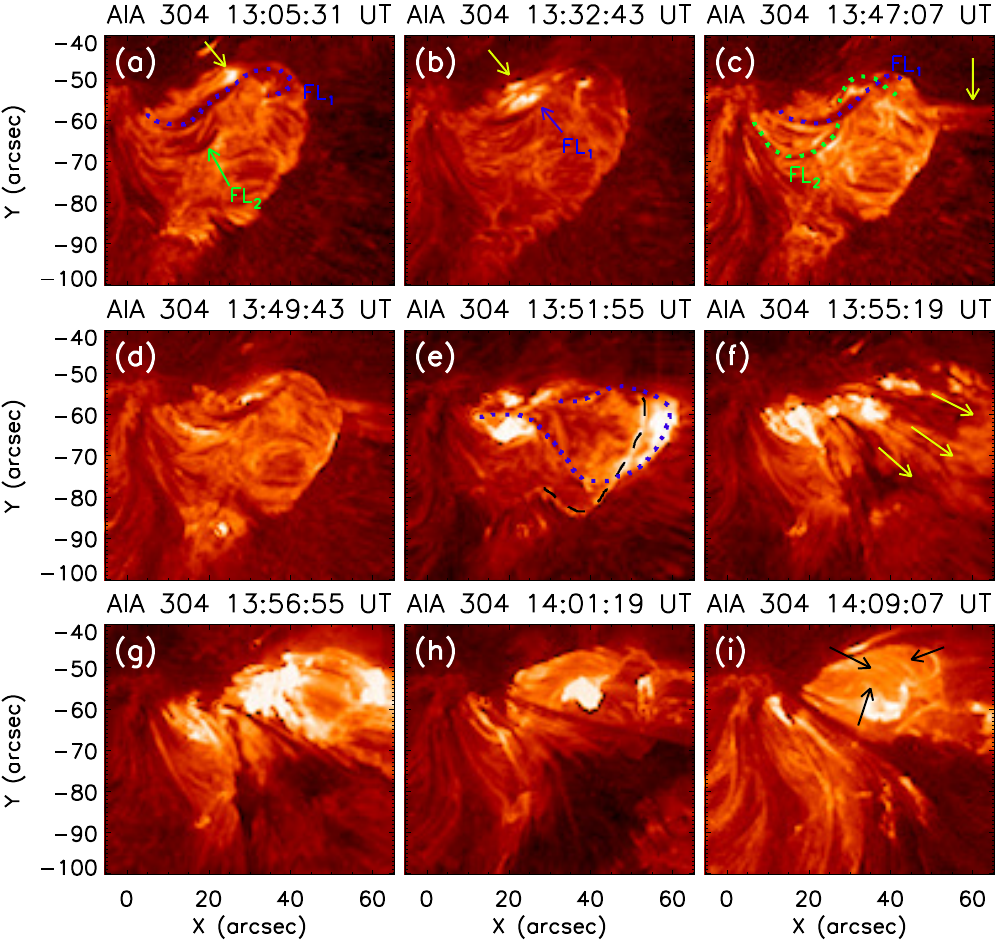}
\caption{Representative AIA 304 \AA\ images depicting the eruption of the filament. The filament FL$_1$ is indicated by the blue dotted curve in panel (a) and blue arrow in panel (b). A second filament (`FL$_2$) is indicated by the green arrow in panel (a). The yellow arrows in panels (a) and (b) indicate small-scale brightenings at the location of the filaments during the pre-flare phase. The blue and green dotted lines in panel (c) indicate the double-decker flux rope configuration formed by FL$_1$ and FL$_2$. The yellow arrow in panel (c) indicate jet-like plasma ejection prior to the onset of the flare. The blue dotted curve in panel (e) indicate erupting filament while the black dashed curve in the same panel indicate a quasi-circular ribbon during the F$_2$ flare. The yellow arrows in panel (f) indicate the direction of the erupting plasma during the flare. The black arrows in panel (i) indicate post-flare arcade. An animation associated with this figure is attached in the online supplementary materials.}
\label{C3.4}
\end{figure*}

Although, the filament situated along the northern edge of the diffused brightening region (FL$_1$), partially erupted during F$_1$, the corresponding PIL showed the presence of a filament channel throughout the lifetime of the diffused brightening region. Before the onset of F$_2$, another filament was observed quite prominently at the same location (i.e., FL$_1$) which is highlighted by a blue dotted line in Figure \ref{C3.4}(a). Importantly, we noticed impressions of a second filament near FL$_1$ which we mark as `FL$_2$' (indicated by the green arrow in Figures \ref{C3.4}(a)). Similar to the pre-flare phase prior to F$_1$, the localised region associated with the filaments, underwent small-scale brightenings during the pre-flare phase of F$_2$ also. We have indicated few such episodes of brightenings by the yellow arrows in Figures \ref{C3.4}(a) and (b). With time, FL$_1$ became more prominent, which is indicated by the blue arrow in Figure \ref{C3.4}(b). Few minutes prior to the onset of F$_2$, the filament FL$_2$ also became very prominent, making a double decker flux rope system in association with FL$_1$. In Figure \ref{C3.4}(c), we indicate the two filaments of the double-decker system by blue and green dotted lines. Notably, this double-decker system is confirmed by the presence of two intertwined flux ropes (Section \ref{fieldlines}). Interestingly, the north-western end of the FL$_2$ was associated with narrow collimated plasma eruption even before the flare was initiated (indicated by the yellow arrow in Figure \ref{C3.4}(c)). Our observations suggest that the flare onset took place as the filament FL$_1$ went through a complete eruption at $\approx$13:51 UT. In Figure \ref{C3.4}(e), we have outlined the erupting FL$_1$ by a blue dotted line. At the same time, a part of the western boundary of the diffused brightening region became very bright implying the formation of a circular ribbon which we have highlighted by the black dashed curve in Figure \ref{C3.4}(e). Soon after the onset of the eruption, the active region was partially masked by the cool erupting plasma (the direction of erupting plasma is indicated by yellow arrows in Figure \ref{C3.4}(f)). During the gradual phase, we observed a dense post flare arcade (indicated by black arrows in Figure \ref{C3.4}(i)) which is expected.

\section{Coronal Magnetic Field Modelling} \label{extrapolation}

\subsection{Extrapolation set up} \label{NLFFF_set_up}
In order to investigate the coronal magnetic configurations prior to the onset of the quasi-circular ribbon flares, we employed a non-linear force free field (NLFFF) extrapolation method with a particular focus to the magnetic atoll (i.e., flaring) region, using the vector magnetograms from the `\textit{hmi.sharp\_cea\_720s}' series at four times: 09:22 UT (prior to F$_1$; Figure \ref{nlfff}), 13:46 UT (prior to F$_2$; Figure \ref{nlfff_c}), 17:34 UT (prior to F$_3$; Figure \ref{nlfff_f3_f4}(a)--(d1)) and 18:58 UT (prior to F$_4$; Figure \ref{nlfff_f3_f4}(e)--(h1)) on 2014 February 16 as the input boundary conditions. Since the photosphere is not force-free, the photospheric magnetograms used as the input boundary conditions were preprocessed, as explained in \citet{Wiegelmann2006}. The optimization based NLFFF code, including the part of preprocessing of input magnetic field, allows a number of `free parameters' for the user's consideration, which are $\nu$, $w_{los}$, $w_{trans}$, $\mu_1$, $\mu_2$, $\mu_3$ and $\mu_4$ \citep[for a quick summary, see][]{Mitra2020b}. In all the four instances of coronal magnetic field extrapolation conducted in this work, we used the following values of the free parameters:

$\nu=0.01; ~~~~~~~~~~~ w_{los}=1; ~~~~~~~~~~~ w_{trans}=\frac{B_{trans}}{max(B_{trans})};\\
~~~~~~~~~\mu_1=\mu_2=1; ~~~~~\mu_3=0.001;~~~~~~~ \mu_4=0.01$.

\noindent Theoretically, in the NLFFF model, the angle between current ($\vec{J}$) and magnetic field ($\vec{B}$) should be 0. However, since the NLFFF-code uses real measurements of magnetic field, small but non-zero values of |$\vec{J}\times\vec{B}$| is expected from the reconstructed magnetic field. Therefore, to assess the quality of the coronal magnetic field reconstruction, the average value of fractional flux ratio ($<|f_i|>=<|(\vec{\nabla}\cdot\vec{B})_i|/(6|\vec{B}|_i/\bigtriangleup x)>$), weighted angle ($\theta_J$) between $\vec{J}$ and $\vec{B}$ can be considered \citep[see,][]{DeRosa2015}. In general, NLFFF solutions returning the values $|\vec{J}\times\vec{B}|\lesssim10^{-2}$, $<|f_i|>\lesssim2\times10^{-3}$, $\theta_J\lesssim10^{\circ}$ are considered as good solutions \citep[see e.g.,][]{DeRosa2015, Thalmann2019}. In Table \ref{table2}, we have listed the values of these parameters corresponding to all the four extrapolations used in this paper. Here, it should be noted that, the extrapolations were conducted over the entire active region (Figure \ref{ar_flux}(a)), however, in this section and in Figures \ref{nlfff}--\ref{nlfff_f3_f4}, we have only shown and discussed the modelled coronal configuration associated with the atoll region.

\begin{table*}
\centering
\caption{Summary of the parameters for assessing the quality of NLFFF extrapolation}
\label{table2}
\begin{tabular}{cccc}
\hline
\hline
Time of extrapolation&<|$\vec{J}\times\vec{B}$|>&<|$f_i$|>&$\theta_J$\\
\hline
09:22 UT&4.16$\times$10$^{-3}$&6.79$\times$10$^{-4}$&8.35$^\circ$\\
13:46 UT&4.82$\times$10$^{-3}$&7.26$\times$10$^{-4}$&8.24$^\circ$\\
17:34 UT&3.89$\times$10$^{-3}$&6.93$\times$10$^{-4}$&7.88$^\circ$\\
18:58 UT&5.67$\times$10$^{-3}$&6.58$\times$10$^{-4}$&7.59$^\circ$\\
\hline
\end{tabular}
\end{table*}

\subsection{Preflare coronal magnetic connectivities} \label{fieldlines}
\begin{figure*}
\includegraphics[width=\textwidth]{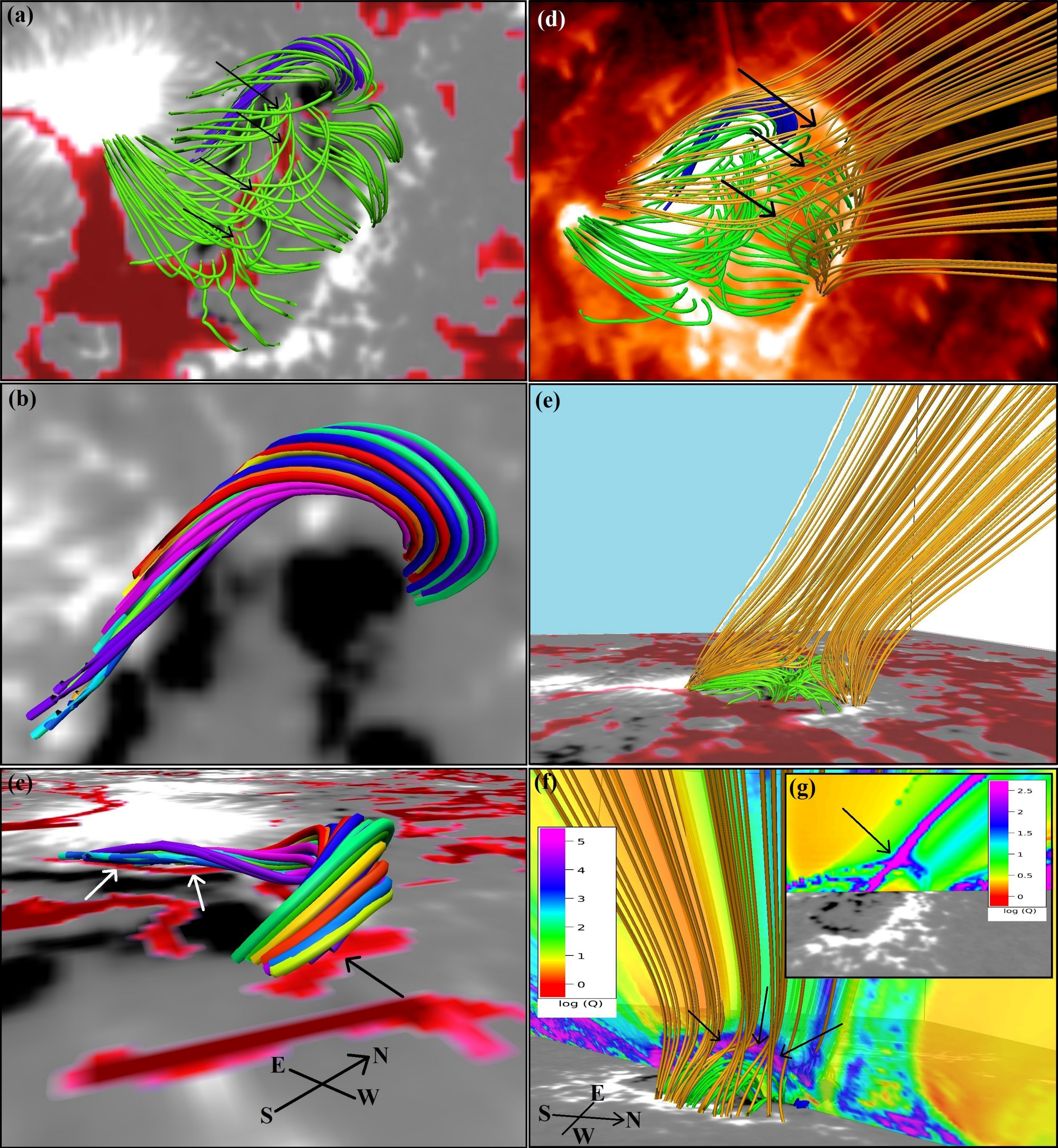}
\caption{Non-linear force free field extrapolation results of the magnetic atoll region prior to the flare F$_1$ (at 09:22 UT) showing the presence of a flux rope and closed magnetic loops within the region (green lines) besides large open field lines originating from the outer magnetic patches of the atoll region (yellow lines). To understand the structures of the flux ropes clearly, we provide zoomed views of the flux rope with multiple colours in panels (b) (top view) and (c) (side view). The red patches over the background HMI magnetogram in panels (a), (c), and (e) represent regions with high $Q$-value ($log(Q)>2$). The arrows in panel (c) highlight the close association between the legs of the flux rope and photospheric regions with high $Q$-value. The Y-Z tilted vertical surface in panel (f) drawn along the yellow lines represent the distribution of $Q$-values. In panel (g), the distribution of $Q$ is shown along a plane passing across the magnetic atoll region i.e., along X-Z plane. The arrow in panel (g) indicate an `X'-shape formed by the regions of high $Q$. For reference, we have included the colour-table showing the distribution of $log(Q)$ values in the box in panels (f) and (g). Top boundary in panels (a), (b), (d) as well as the sky-coloured boundary in panel (e) represent north. We have plotted a compass for representing the direction in panels (c) and (f). An AIA 304 \AA\ image of 09:26 UT has been plotted in the background in panel (d).}
\label{nlfff}
\end{figure*}

\begin{figure*}
\includegraphics[width=\textwidth]{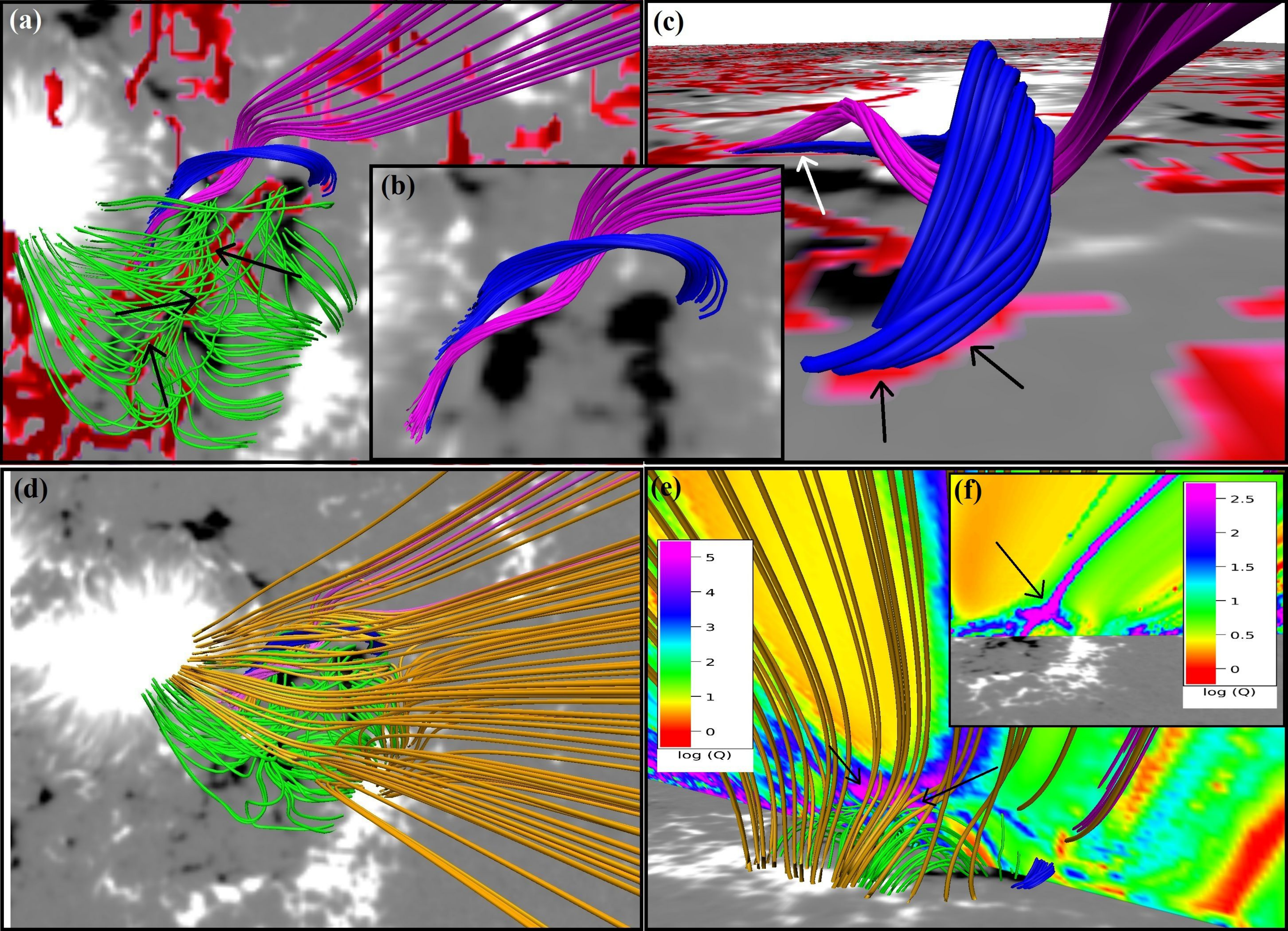}
\caption{Non-linear force free field extrapolation results on the magnetic atoll region prior to the F$_2$ flare (at 13:46 UT) showing the presence of a flux rope (blue lines) and closed magnetic loops within the region (green lines) as well as large open field lines originating from magnetic patches of the atoll region (yellow lines). A set of twisted field lines having origin in the south-western end of the polarity inversion line and extending north-westward as a part of open field lines i.e., `open flux rope' are shown in pink colour. In panels (b) and (c), only the two flux ropes are shown from top and side views, respectively. The red patches over the background HMI magnetogram in panels (a) and (c) represent regions with high Q-value ($log(Q)>2$). The arrows in panel (c) highlight the close association between the legs of the flux ropes and photospheric regions with high $Q$. Panels (e) and (f) are same as Figures \ref{nlfff}(f) and (g), respectively. Top boundary in panels (a), (b) and (d) represent north. Top boundary in panel (c) represents east.}
\label{nlfff_c}
\end{figure*}

NLFFF extrapolation results at 09:22 UT readily indicates the presence of a flux rope in the apparent location of FL$_1$ which are drawn by blue lines in Figures \ref{nlfff}(a) and (d). For a better understanding of the structure of the flux rope, we have shown only the field lines constituting the flux rope from top and side views in Figure \ref{nlfff}(b) and (c), respectively, where different field lines are plotted in different colours.  The flux rope is enveloped by a set of highly sheared overlying closed loops (shown by the green lines in Figure \ref{nlfff}). Notably, these field lines connect the outer positive polarity regions to the central dispersed negative polarity regions. The green lines are surrounded by a set of open field lines (shown by yellow color in Figures \ref{nlfff}(d)--(f)) which originate from the outer positive polarity regions. Interestingly, the boundary of the diffused, nearly-circular flare brightening apparently delineates the footpoints of the open field lines (cf. the modelled yellow lines and the background AIA 304 \AA\ image in Figure \ref{nlfff}(d)). In this way, the entire structure made of the green and yellow lines resembles a spine-fan-like configuration.However, although few low-lying null-points were located close to the bottom boundary of the extrapolation volume, we could not find presence of any null-point near or within our region of interest, suggesting that the coronal configuration associated with the flaring activities reported in this article, differed from the 3D spine-fan configuration. Further, the laterally extended nature of the spine-like lines (indicated by the black arrows in Figures \ref{nlfff}(a) and (d)) is also uncharacteristic of the spine-fan configuration. In Figure \ref{nlfff}(e), we show the whole configuration from a different angle for an overall visualisation. This atypical, spine-fan-like configuration is also recognised prior to the onset of F$_2$ which is indicated by the black arrows in Figures \ref{nlfff_c}(d)--(e)). Our extrapolation results reveal that this spine-fan-like configuration decays significantly in spatial extent afterwards (Figures \ref{nlfff_f3_f4}(a) and (e)) vis-\`{a}-vis changes in the corresponding photospheric magnetic field structure of the magnetic atoll region (Figures \ref{allflares}(j)--(m)).

The model field structure during the preflare phase of F$_2$ reveals two sets of twisted field lines intertwined with each other in a double-decker flux rope configuration (shown by blue and pink colours in Figures \ref{nlfff_c}(a)--(c)). While the twisted field lines shown in blue are similar to the flux rope identified during the pre-flare phase of F$_1$ (i.e., FL$_1$), the apparent twist associated with it is found to be more than the previous flux rope (see Table \ref{table3}). The other set of twisted field lines shown in pink colour is rather interesting. One leg of these lines is situated in the PIL region where one leg of the blue flux rope is also located. Further, we note that, the location of the pink lines is same as the apparent location of FL$_2$ (cf. Figures \ref{nlfff_c}(b) and \ref{C3.4}(c)). However, while the blue flux rope is anchored to the photosphere at both the ends, the field lines constituting the pink flux rope is anchored only at its southern end. At the northern end, the field lines of the flux rope become a part of the open lines instead of terminating on the photosphere (cf. the open end of the pink lines in Figures \ref{nlfff_c}(a) and the open yellow lines in Figure \ref{nlfff_c}(d)). In Figures \ref{nlfff_c}(b) and (c), we show only the double-decker flux rope configuration from top and side views for a better understanding of their structures. In Figure \ref{nlfff_f3_f4}, we show the modelled magnetic configuration above the magnetic atoll region prior to the onset of the F$_3$ (Figures \ref{nlfff_f3_f4}(a)--(d1)) and F$_4$ (Figures \ref{nlfff_f3_f4}(e)--(h1)) flares. We find that, similar to F$_1$, prior to the F$_3$ and F$_4$ flares also, single flux rope structures are identified from the modelled coronal magnetic field which are shown from top views in Figures \ref{nlfff_f3_f4}(a), (b) and \ref{nlfff_f3_f4}(e), (f), respectively, and side views in Figures \ref{nlfff_f3_f4}(c) and (g), respectively.

\subsection{Distribution of Squashing factor ($Q$)} \label{qsl}
\begin{figure*}
\includegraphics[width=\textwidth]{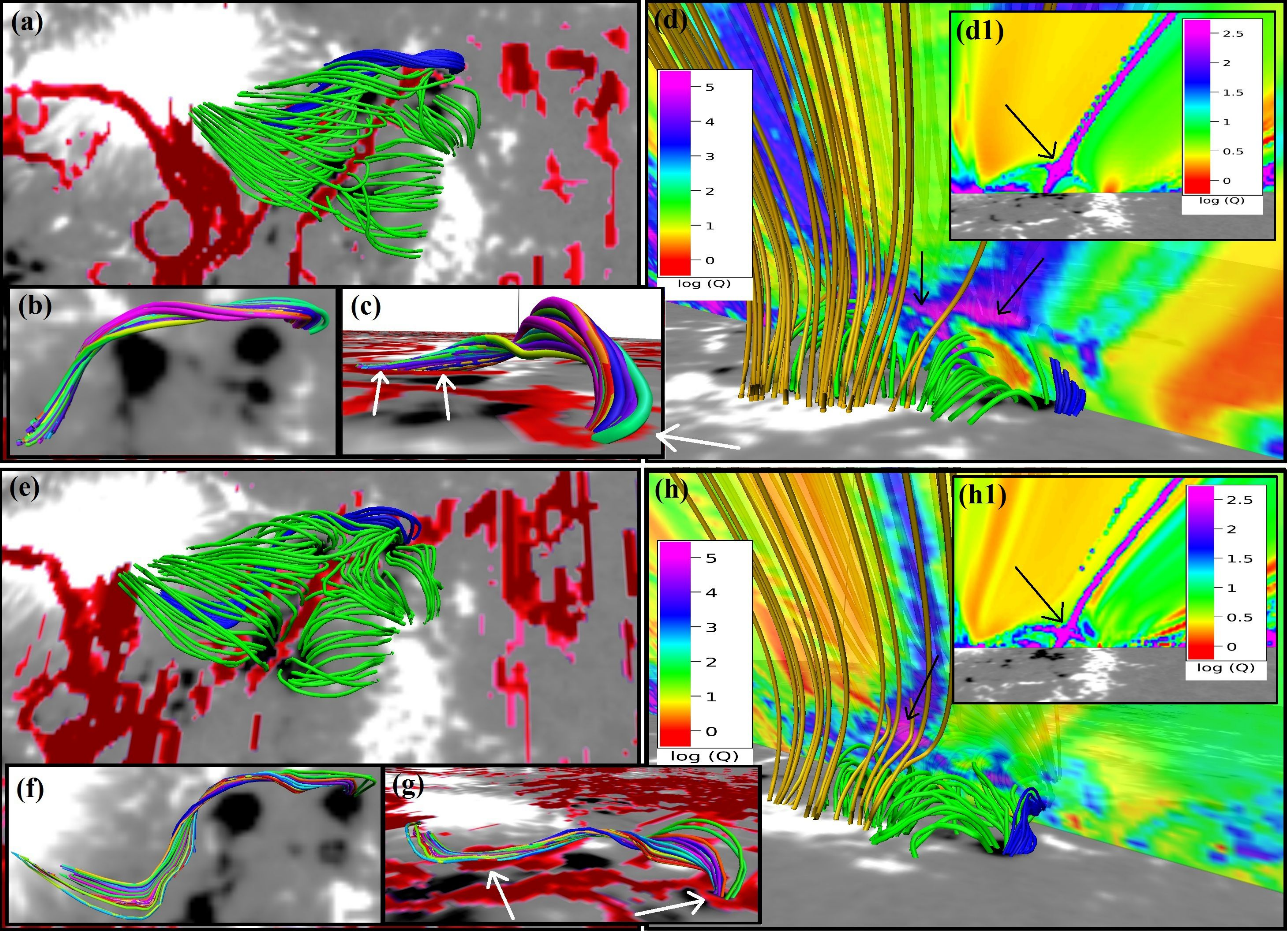}
\caption{NLFFF extrapolation results showing the coronal configurations prior to the F$_3$ (at 17:34 UT; panels (a)--(d1)) and F$_4$ flare (at 18:58 UT; panels (e)--(h1). Panels (a) and (e) show the flux ropes (in blue colour) and the inner fan-like lines (in green colour) from top view. The flux ropes prior to the F$_3$ and F$_4$ flares are exclusively shown in multiple colours from the top view in panels (b) and (f), respectively, and from side views in panels (c) and (g), respectively. Panels (d)--(d1) and (h)--(h1) are same as Figures \ref{nlfff}(f)--(g), respectively. The arrows in panels (c) and (g) highlight the close association between the legs of the flux ropes and photospheric regions with high $Q$.}
\label{nlfff_f3_f4}
\end{figure*}

NLFFF modelling reveals a complex coronal magnetic configuration prior to the successive four flares. While the overall coronal structures resemble a fan-spine-like configuration, they lack a coronal null-point. To further investigate the fan-spine-like configuration, we calculated the squashing factor ($Q$) in the active region using the NLFFF extrapolated magnetic fields in the extrapolation volume. The relevant panels in Figures \ref{nlfff}--\ref{nlfff_f3_f4} display the photospheric regions associated with high $Q$ ($log(Q)>2$) values by the red coloured patches. From Figures \ref{nlfff}(a), \ref{nlfff_c}(a), \ref{nlfff_f3_f4}(a) and (e), we readily observe that along the elongated footpoint locations of the green lines over the negative polarity magnetic field region (indicated by the black arrows in Figures \ref{nlfff}(a) and \ref{nlfff_c}(a)), the $Q$-values are higher than 10$^2$ which provides substantial evidence for the laterally extended nature of the spine-like lines. Notably, the footpoint regions of the flux ropes are also found to be associated with high $Q$-values which can be inferred from the red coloured patches indicated by the arrows in Figures \ref{nlfff}(c), \ref{nlfff_c}(c), \ref{nlfff_f3_f4}(c) and (g).

To understand the variation of $Q$-values along the laterally extended spine-like lines, we draw tilted vertical planes passing through the outer spine-like lines (i.e., the yellow lines shown in Figures \ref{nlfff}(d)--(e), \ref{nlfff_c}(d)); distribution of $Q$ along which are shown in Figure \ref{nlfff}(f), \ref{nlfff_c}(e) and \ref{nlfff_f3_f4}(d) and (h). We find that immediately over the inner fan-like lines (shown by green color), the $Q$-values approached maximum values ($log(Q)\gtrsim$5) signifying drastic change in the magnetic connectivity between the green and the yellow lines. The extended arc-shaped high $Q$ region (shown in purple color and the arrows in Figures \ref{nlfff}(f), \ref{nlfff_c}(e) \ref{nlfff_f3_f4}(d) and (h)) in the absence of coronal nulls, suggests the presence of an HFT between the green and yellow lines. In Figure \ref{nlfff}(g), we plot $Q$-values along a plane that crosses perpendicularly the tilted plane of Figure \ref{nlfff}(f), i.e., it shows the variation of $Q$ across the spine-fan-like configuration. From this panel, we readily observe the `X'-shape formed by high $Q$-values (indicated by the black arrow) which further confirms the presence of the HFT in the coronal magnetic configuration above the atoll region. Notably, similar configurations are also found prior to the onset of the subsequent flares (indicated by the arrow in Figures \ref{nlfff_c}(f), \ref{nlfff_f3_f4}(d1) and (h1)); however, with the decay of the magnetic atoll region, the extended coronal region of high $Q$-values over the inner fan-like lines (indicated by arrows in Figures \ref{nlfff}(f), \ref{nlfff_c}(e) and \ref{nlfff_f3_f4}(d)), gradually reduced and concentrated to a point-like structure prior to the onset of the F$_4$ flare (indicated by the black arrow in Figure \ref{nlfff_f3_f4}(h)).

\subsection{Calculation of twist number and magnetic decay index} \label{sec_twist}
Both GONG H$\alpha$ and AIA 304 \AA\ images suggest the filaments going through eruptive evolution during the flares (Figures \ref{halpha}, \ref{M1.1}, \ref{C3.4}). The presence of the filaments are confirmed by the presence of the flux ropes identified in the NLFFF-modelled coronal configuration (Figures \ref{nlfff}, \ref{nlfff_c}, \ref{nlfff_f3_f4}). For a quantitative assessment of the twists of the flux ropes, we compare distribution of twist number \citep[$T_w$; see,][]{Berger2006} associated with the location of the flux ropes within the extrapolation volume, defined as 
\begin{equation} \label{eq_tw}
T_w=\frac{1}{4\pi}\int_{L}\frac{(\nabla\times\vec{B})\cdot\vec{B}}{B^2} dl
\end{equation}
where $L$ is the length of the flux rope. In Table \ref{table3}, we present average twist number ($T_w$) associated with the flux ropes corresponding to consecutive flares prior to their onset (F$_1$--F$_4$). We find that $T_w$ increased successively from F$_1$ to F$_4$. While $T_w$ prior to F$_1$ was only $\lesssim$0.93, it increased to $\approx$1.22 prior to the onset of F$_4$.

\begin{figure*}
\includegraphics[width=\textwidth]{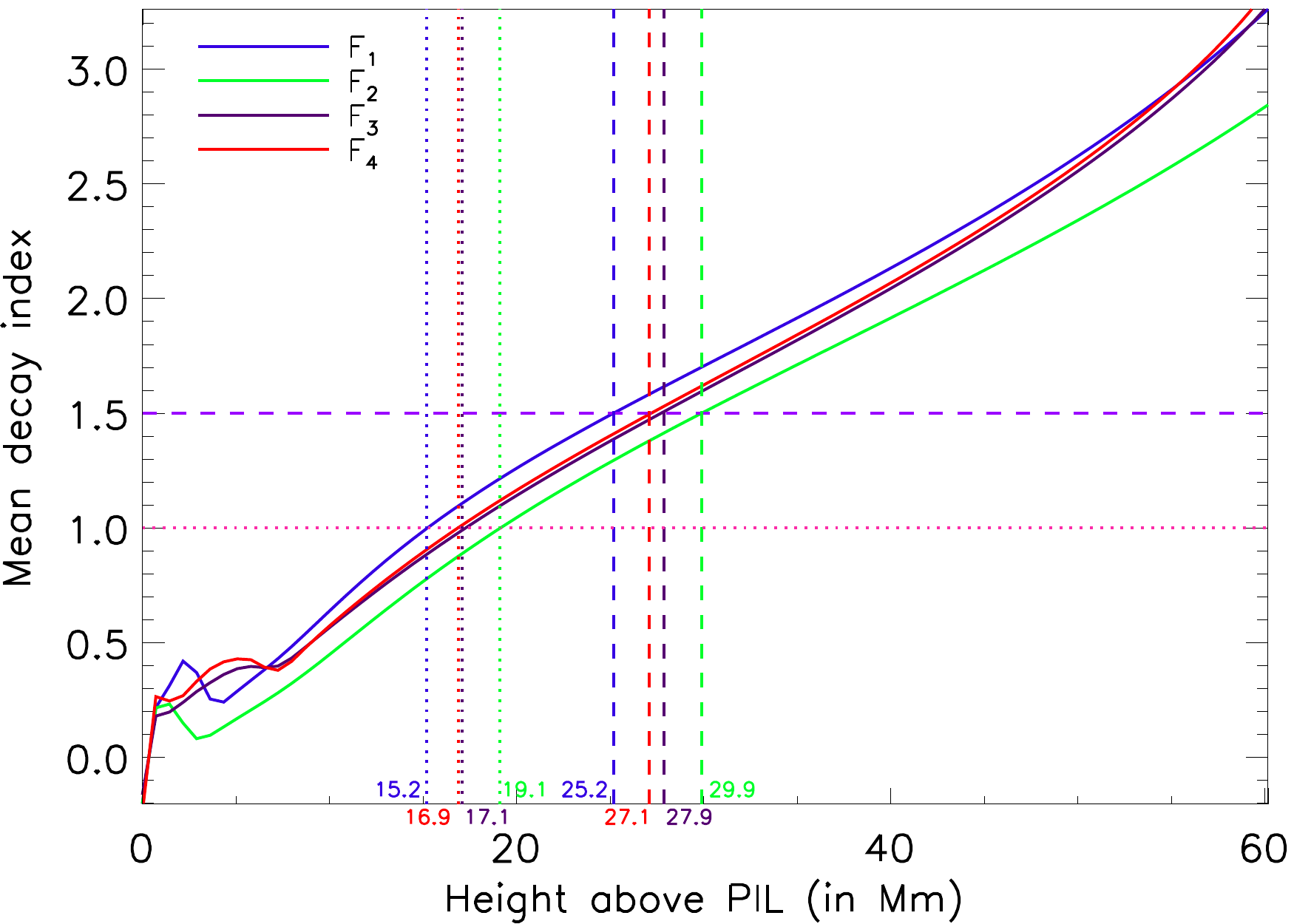}
\caption{Plot of decay index ($n$) above the PIL as a function of height, prior to the four events. The values of decay index $n=1.0$ and 1.5 are indicated by the pink dotted and purple dashed horizontal lines. The critical heights corresponding to the values $n=1.0$ and 1.5 prior to the flares are represented by the dotted and dashed vertical lines, respectively, and noted at the bottom of the plot. Colours of these vertical lines and the values of critical heights corresponding to the events are synchronised with the colours of the decay index plots.}
\label{decay_plot}
\end{figure*}

To understand how the horizontal magnetic field changed with height over the flux ropes, we calculate magnetic decay index ($n$). For this purpose, we have considered the PILs over which the flux ropes were situated and computed average decay index along vertical surfaces above it. In Figure \ref{decay_plot}, we plot the variation of magnetic decay index averaged over the PILs, with height; where we have indicated the critical heights at which the value of decay index reached $n=1.0$ ($h_{crit}(n=1.0)$) and $n=1.5$ ($h_{crit}(n=1.5)$) prior to all the four flares, by the dotted and dashed vertical lines. In Table \ref{table3}, we have summarised the values of $h_{crit}$ and approximate maximum heights of the flux ropes before the onset of all the four flares. Our calculations suggest that prior to the F$_1$ flare, $h_{crit}$ was $\approx$25 Mm while it slightly increased to $\approx$27--30 Mm prior to the subsequent flares. However, the maximum heights of the flux ropes prior to the onset of the flares were found to be only $\approx$4--5 Mm (Table \ref{table3}) which are much less compared to the critical heights.

\begin{table*}
\centering
\caption{Summary of the twist number $|T_w|$ and critical height ($h_{crit}$) for magnetic decay index $n=1.0$ and 1.5, prior to the four flares from the magnetic atoll region}
\label{table3}
\begin{tabular}{ccccc}
\hline
\hline
Flare&$|T_w|$&$h_{crit}(n=1.0)$ (Mm)&$h_{crit}(n=1.5)$ (Mm)& Maximum height of\\
Id.&&&& the flux rope (Mm)\\
\hline
F$_1$&$\approx$0.93$\pm$0.11&$\approx$15&$\approx$25&$\approx$ 4\\
F$_2$&$\approx$1.12$\pm$0.18&$\approx$19&$\approx$30&$\approx$ 4\\
F$_3$&$\approx$1.20$\pm$0.20&$\approx$17&$\approx$28&$\approx$ 5\\
F$_4$&$\approx$1.22$\pm$0.20&$\approx$17&$\approx$27&$\approx$ 4\\
\hline
\end{tabular}
\end{table*}

\section{Discussion} \label{dscsn}
In this article, we present a detailed analysis of four successive flares (F$_1$, F$_2$, F$_3$, F$_4$; in the order of their occurrence) associated with quasi-circular ribbons. These flares were originated from the active region NOAA 11977 on a single day within a period of $\approx$11 hour (see Table \ref{table1}). The flaring region possessed a fan-spine-like configuration that involved a hyperbolic flux tube (HFT). Observationally, the fan-spine-like configuration was identified in the form of a region with diffused EUV brightening that spread within a circular base prior to F$_1$. The fan-spine-like structure decayed following each flare and, as a consequence, the prominent quasi-circular brightening identified prior to F$_1$ also decomposed and simplified during the succeeding flares.

Our observations revealed that the development of the interesting coronal configuration prior to F$_1$, facilitating the origin of the subsequent morphologically similar flaring events, was eventually connected with the emergence of negative polarity patches on the photosphere that formed a magnetic atoll region (Figure \ref{dome_form}). Further, the location over which negative polarity patches emerged can be characterised by a geometric stadium shape\footnote{\url{http://mathworld.wolfram.com/Stadium.html}} with the longitudinal dimension lying along northwest-southeast (NW-SE) direction (Figure \ref{dome_form}(f)). These negative polarity regions were surrounded by positive polarity regions on northeast (NE) and southwest (SW) sides. Thus, the photospheric structure of the flaring region consisted of a longitudinally stretched magnetic patch bounded by regions of opposite polarity on both sides. NLFFF extrapolation results suggested a complex configuration in which open field lines originating from the positive polarity regions enclosed an extended fan-like structure (Figures \ref{nlfff}--\ref{nlfff_f3_f4}) such that the overall coronal magnetic configuration resembled the topology of pseudo-streamers \citep{Wang2007, Titov2011, Titov2012, Masson2014} albeit in a much smaller spatial scale. Notably, since pseudo-streamers connect coronal holes of the same polarity, involving even number of PILs, two-dimensional depiction of the cross-section of pseudo-streamers indicate the presence of X-shaped high Q-structures \citep{Titov2012}, which may represent true 3D null-points or topological structures such as separators connecting multiple null-points, HFTs etc. \citep{Gibson2017}. In the absence of any null-point, the coronal magnetic configuration, derived from our analysis, can be physically well interpreted by considering a small-scale pseudo-streamer involving an HFT.

The presence of the HFT in the flaring region was further confirmed by their cross-sectional X-shaped high $Q$-regions (Figures \ref{nlfff}(g), \ref{nlfff_c}(f), \ref{nlfff_f3_f4}(d1) and (h1)). The high $Q$ values of these regions imply high gradient of magnetic field (i.e., QSLs) within the diffused brightening region. Intense current sheets are formed naturally around QSLs as gradient in magnetic field contributes toward the generation of current \citep{Lau1990, Priest1996, Aulanier2005, Demoulin2007}. Using MHD simulations, formation of electric current has also been demonstrated in magnetic flux ropes, coronal sigmoids, beneath an erupting flux rope etc. \citep[see, e.g.,][]{Wilmot2009, Pariat2009, Aulanier2009}. Joule heating due to dissipation of these currents associated with the QSL formed by the inner fan-like lines was most likely responsible for the diffused EUV brightness confined within a quasi-circular border and was most prominent during the pre-flare phase of F$_1$ (see Figure \ref{dome_form}(i)). Evidently, the dome-shaped active pre-flare coronal structure, observed in EUV, was co-spatial with the photospheric magnetic atoll region.

EUV images clearly revealed the formation of distinct quasi-circular flare ribbons along the boundary of dome-shaped pre-flare structure as the filament eruption proceeded (Figures \ref{halpha}, \ref{M1.1} and \ref{C3.4}). Notably, prior to the flares, the filament resided within the EUV-dome, i.e., modelled fan-spine-like configuration (Figures \ref{nlfff}--\ref{nlfff_f3_f4}). This is definitive signature that the quasi-circular ribbon flares were triggered as the erupting flux rope interacted with the fan-like separatrix surface. Using MHD simulations, it has been established that stressed QSL regions can give rise to slipping reconnections even without the presence of a coronal null-point and for sufficiently thin QSLs and high resistivities, the field line footpoints can slip-run at super-Alfv{\'e}nic speeds along the intersection of the QSLs \citep[slip running reconnection;][]{Aulanier2006}. While studying a circular ribbon in association with a coronal null-point topology, \citet{Masson2009} observed that slip-running reconnection and null-point reconnection can occur sequentially. They also found that $Q$ is a highly effective parameter which determines which mode of reconnection will occur in a null-point topology: cut-paste type null-point reconnetion occurs when the value of $Q$ reaches infinity and slipping (or slip-running) reconnection occurs for lesser values of $Q$. As observational signature of slipping/slip-running reconnection, circular ribbon and remote brightening can be highlighted \citep{Masson2009}; while the null-point reconnection usually gives rise to collimated eruptions i.e., coronal jets or H$\alpha$ surges \citep{Pariat2009b, Pariat2010}. 

Notably, we observed collimated eruption of plasma prior to the onset of the C3.4 flare (Figure \ref{C3.4}(c)). AIA 304 \AA\ images clearly revealed two adjacent filaments at the flaring location that apparently crossed each other (Figure \ref{C3.4}(c)). NLFFF extrapolation clearly identified two braided flux ropes within the atoll region (Figure \ref{nlfff_c}). Such arrangement of intertwining flux ropes is called `double-decker flux rope systems' \citep[see e.g., ][]{Liu2012, Cheng2014b, Mitra2020b}. Jet-like plasma ejections resulting from the interaction between the flux ropes within a double-decker system has been reported in \citet{Mitra2020b}. Further, the double-decker region reported in \citet{Mitra2020b} was associated with a set of open field lines which guided the eruption of collimated jets. NLFFF model magnetic field structure prior to the onset of F$_2$ revealed that one end of one flux rope within the double-decker system was directly connected to the open spine-like-lines (i.e., open flux rope; shown in the pink colour in Figure \ref{nlfff_c}). These findings led us to conclude that the jet-like eruption was triggered as a result of the interaction between the two flux ropes of the double-decker flux ropes system; while, the open field lines was responsible for guiding eruption in a collimated manner. We also clarify that, in our case, the collimated eruption observed prior to the C3.4 flare should not be related with magnetic reconnection at the HFT. Here, it is worth mentioning that magnetic structures similar to the open flux rope shown in Figure \ref{nlfff_c}, where magnetic field lines constituting the flux rope, becomes open at one end, has been previously noted by \citet{Lugaz2011, Janvier2016}. While numerically investigating evolution of eruptive flares from complex photospheric configurations of the active regions NOAA 10798 and 11283, respectively, both the studies found that opening of the field lines of flux rope structures can lead to and guide solar eruptions resulting in the formation of CMEs.

We would like to highlight the association of flux ropes with QSL which we noted in all the four cases (Figures \ref{nlfff}(c), \ref{nlfff_c}(c), \ref{nlfff_f3_f4}(c) and (g)). Our analysis readily revealed that the photospheric regions associated with the legs of the flux ropes were characterised by high $Q$-values. High $Q$-values at the boundaries of flux ropes essentially associated with two different sets of magnetic field lines: one set forming the flux rope and the other set forming the relatively potential, ambient magnetic field \citep[see e.g.,][]{Savcheva2012, Zhao2016, Janvier2016, Guo2019}.

The atoll region produced four homologous flares on 2014 February 16 (Table \ref{table1}). All the flares triggered as a filament lying over the PIL between the positive polarity sunspot and negative polarity regions, got destabilised. The erupting filament interacted with the fan-spine configuration which led to some degree of circular ribbon brightening during all the flares (Figure \ref{allflares}). Prior to the onset of each flare, we could identify instances of flux cancellation from the PIL region (Figure \ref{ar_flux}) as well as localised brightenings beneath the filaments (Figure \ref{M1.1} and \ref{C3.4}). These observational findings support the tether-cutting model of solar eruptions \citep{Moore1992, Moore2001}. We also explored the possibility of torus and kink instabilities as the triggering mechanism, by computing the decay index and twist numbers. Our analysis suggests that the critical decay height for $n=$1.5 above the PIL of the flux rope prior to all the four flares remained consistent within the heights of $\approx$25--30 Mm (Table \ref{table3}). Statistical surveys concerning the torus instability as the triggering mechanism for eruptive flares by \citet{WangD2017, Baumgartner2018} revealed the ciritical decay height to lie within the ranges of $\approx$36$\pm$17 Mm and 21$\pm$10 Mm. Our results of decay height calculations are in well agreement with these statistically established values, suggesting a favourable role played by torus instability. However, considering the average decay index over the PIL, as followed in the present analysis, is rather a simplistic approach. Detailed studies devoted to the analysis of decay index \citep[see e.g.,][]{Zuccarello2014, Zuccarello2017, Myshyakov2020} have shown that a flux rope eruption can take place even when the critical height (h$_{crit}$) becomes sufficiently low in a few discrete locations over the flux rope. Further, theoretical studies have shown that the critical height strongly depend on the magnetic topology \citep{Kliem2014}. In this context we note that, the coronal magnetic configuration associated with all the four flares reported in this paper was much complex compared to the general cases without spine-fan configurations. Rapid decay of magnetic field within the fan-surface is expected which in turns results in high values of magnetic decay index. Average twist number associated with the flux ropes increased successively from $\approx$0.93 prior to F$_1$ to $\approx$1.22 prior to $F_4$ (Table \ref{table3}). Although, the increase in twist suggest higher storage of magnetic free energy in the flux ropes, the critical value of twist number for resulting in the destabilisation of the flux ropes was statistically established to be $|T_w|\approx$2 \citep{Duan2019}. Therefore, we could not find any conclusive evidence for kink instability as the triggering mechanism for the homologous flares reported in this paper.

It is also noteworthy that, although a clear circular ribbon appeared during the M-class flare (F$_1$), no observable signature of null-point-like reconnection (i.e., jet/surge, breakout type eruption etc.) was observed during it. AIA 304 \AA\ images during the flare clearly suggested that the erupting filament experienced an apparent sliding motion from the northern end to the southern end within the diffused brightening region. Plasma eruption from one end of the filament was observed only when the sliding filament eventually reached the southern boundary of the diffused brightening region. These observational features suggest the occurrence of predominantly QSL-reconnection during the M1.1 flare, rather than the reconnection between the inner and outer fan-like lines. On the other hand, all the subsequent C-class flares (F$_2$--F$_4$) evolved with complete eruption of the filaments from the diffused brightening region, clearly implying reconnection at the HFT. Here we remember that twist number associated with the flux ropes successively increased from F$_1$ to F$_4$. Additionally, negative flux within the magnetic atoll region decreased almost monotonically following the onset of F$_1$ flare signifying less constraining energy stored in the fan-spine configuration during the subsequent flares compared to F$_1$. Therefore, we speculate that excess energy stored in the flux rope (in the form of higher twist number) and less energy stored in the constraining magnetic field resulted in the complete destruction of the fan-spine-like configuration during the subsequent C-class flares; while, during the M1.1 flare, less twisted flux rope did not have enough energy to trigger exchange type reconnection at the HFT. 

In summary, all the eruptive flares initiated from a diffused brightening region which formed over a complex magnetic configuration where dispersed negative polarity regions were surrounded by positive polarity regions (magnetic atoll region). The coronal configuration associated with the magnetic atoll region was a fan-spine-like configuration that involved an HFT situated in the corona above the elongated parasitic negative polarity regions; a configuration similar to those of pseudo-streamers, in a much smaller spatial scale. All the four flares were initiated as a flux rope was activated and erupted within the fan dome. Prior to all the flares, we observed localised brightenings associated with the filaments as well as small-scale flux cancellation from the PIL region which supports the tether-cutting model of solar eruption. The magnetic decay index reached to the value of 1.5 within low coronal heights prior to all the four flares signifying favourable coronal conditions for driving successful eruption of the flux ropes. Interaction between the erupting flux rope and the fan-like-separatrix surface gave rise to circular ribbon brightening during the flares. During the first flare, the erupting flux rope with relatively less twist, could not trigger reconnection in the HFT; while, during the subsequent flares, the flux ropes having relatively higher twist, could blow out the already decaying fan-spine-like configuration leading to the complete eruption of the core fields. We further emphasize that, occurrence of successive quasi-circular ribbon flares from complex fan-spine-like configurations including HFTs, have been rarely reported in the literature and subsequent studies involving theoretical and observational analyses of similar events are essential to reach to a general understanding of the complex coronal configurations in the solar atmosphere.

%%%%%%%%%%%%%%%%%%%%%%%%%%%%%%%%%%%%%%%%%%%%%%%%%%%%%%%%%%%%%%%%%%%%%%%%%%%

\section*{Acknowledgements}
\addcontentsline{toc}{section}{Acknowledgements}

We would like to thank the \textit{SDO} team for their open data policy. \textit{SDO} is NASA's mission under the Living With a Star (LWS) program. This work utilises GONG data from NSO, which is operated by AURA under a cooperative agreement with NSF and with additional financial support from NOAA, NASA, and USAF. We are thankful to Dr. Thomas Wiegelmann for providing the NLFFF code. We are also thankful to the anonymous referee for his/her important comments and suggestions which enhanced the scientific content and overall presentation of the article.

\section*{Data Availability}
\addcontentsline{toc}{section}{Data Availability}

Observational data from AIA and HMI on board \textit{SDO} utilised in this article are available at \url{http://jsoc.stanford.edu/ajax/lookdata.html}. GONG H$\alpha$ data used in this article are available at GONG data archive (\url{https://gong2.nso.edu/archive/patch.pl?menutype=a}). The NLFFF code employed in this article for coronal magnetic field modelling is provided by Dr. Thomas Wiegelmann. Different aspects of the code are explained and discussed in \url{https://doi.org/10.1023/B:SOLA.0000021799.39465.36}, \url{https://doi.org/10.1007/s11207-006-2092-z}, \url{https://doi.org/10.1051/0004-6361/201014391}, \url{https://doi.org/10.1007/s11207-012-9966-z}. The IDL-based code used for the computation of $Q$ and $T_w$ is available at \url{http://staff.ustc.edu.cn/~rliu/qfactor.html}.

\section*{Supplementary Material}
\addcontentsline{toc}{section}{Supplementary Material}

Videos are attached with Figures \ref{allflares}, \ref{M1.1} and \ref{C3.4}, which are available in the online article.\\

\addcontentsline{toc}{section}{References}
\bibliographystyle{mnras}

\end{document}